\documentclass[pra,twocolumn,10pt,aps,superscriptaddress,longbibliography]{revtex4-2}  % for review and submission
\setcitestyle{square}
\usepackage{graphicx}  % needed for figures
\usepackage{float}
\usepackage{dcolumn}   % needed for some tables
\usepackage{bm}        % for math
\usepackage{amssymb}   % for math
\usepackage{amsmath}
\usepackage{isomath}
\usepackage[usenames, dvipsnames]{color}
\usepackage{soul}
\usepackage{physics}
\usepackage[normalem]{ulem}
\usepackage{xpatch}
\usepackage{IEEEtrantools}
\usepackage{subcaption}
\usepackage{pgffor}
\usepackage{xprintlen}
\usepackage{comment}

\usepackage{xcolor}
\usepackage{hyperref}
\hypersetup{
     colorlinks  = true,
     linkcolor    = red,
     citecolor    = cyan,
     urlcolor     = magenta
     }

\newcommand{\w}{\mathrm{\Omega}}

\begin{document}

\widetext
\title{Dirac Solitons and Topological Edge States in the \\ $\beta$-Fermi-Pasta-Ulam-Tsingou dimer lattice} % Force line breaks with \\

\author{Rajesh Chaunsali}
\email{rchaunsali@iisc.ac.in}
%\affiliation{LAUM, CNRS-UMR 6613, Le Mans Universit\'{e}, Avenue Olivier Messiaen, 72085 Le Mans, France}

\affiliation{Department of Aerospace Engineering, Indian Institute of Science, Bangalore 560012, India}

\author{Panayotis G. Kevrekidis}
\affiliation{Department of Mathematics and Statistics, University of Massachusetts, Amherst, MA 01003-4515, USA}

\author{Dimitri Frantzeskakis}
\affiliation{Department of Physics, National and Kapodistrian University of Athens, Panepistimiopolis, Zografos, Athens 15784, Greece}

\author{Georgios Theocharis}
\email{georgios.theocharis@univ-lemans.fr}
\affiliation{LAUM, CNRS-UMR 6613, Le Mans Universit\'{e}, Avenue Olivier Messiaen, 72085 Le Mans, France}

\date{\today}% It is always \today, today,

\begin{abstract}
We consider a dimer lattice of the Fermi-Pasta-Ulam-Tsingou (FPUT) type, where alternating linear couplings have a controllably small difference, and the cubic nonlinearity ($\beta$-FPUT) is the same for all interaction pairs. We use a weakly nonlinear formal reduction within the lattice bandgap to obtain a continuum, nonlinear Dirac-type system. We derive the Dirac soliton profiles and the model’s conservation laws analytically. We then examine the cases of the semi-infinite and the finite domains and illustrate how the soliton solutions of the bulk problem can be “glued” to the boundaries for different types of boundary conditions. We thus explain the existence of various kinds of nonlinear edge states in the system, of which only one leads to the standard topological edge states observed in the linear limit. We finally examine the stability of bulk and edge states and verify them through direct numerical simulations, in which we observe a soliton-like wave setting into motion due to the instability.
\end{abstract}

\pacs{45.70.-n 05.45.-a 46.40.Cd}% PACS, the Physics and Astronomy
\keywords{}
\maketitle

\section{Introduction}  \label{Section0}

The study of systems of the Fermi-Pasta-Ulam-Tsingou (FPUT) type~\cite{FPU55} has an exciting and long history within nonlinear science~\cite{FPUreview}. More recently, the relevant topics received substantial attention due to experimental connections~\cite{Zab5}. For instance, granular crystals have offered a reasonably mature platform where various nonlinear phenomena are explored, e.g., solitons, discrete breather, and dispersive shock dynamics~\cite{Nester2001, gc_review, yuli_book, granularBook}. Additionally, diverse platforms based on magnets~\cite{Mehrem2017, Chong_2021} and origami cells~\cite{sci_adv} have also been considered.

On the other hand, the exploration of nonlinear partial differential equations (PDEs) of the Dirac type has also recently gained considerable traction. This is due to the emergence of such equations for boson gases confined in honeycomb lattices~\cite{Carr2009,l25} and light propagation in photorefractive honeycomb lattices~\cite{l30,l29};  the latter theme has allowed for the observation of key features such as conical diffraction, among others. These efforts have led to a wide range of mathematical works devoted to studying solitary waves and their stability in such systems~\cite{our_nab,nabil2}.

Furthermore, the third axis of problems with substantial research activity has been on band topology and its potential impact on designing new materials and structures for various engineering applications. Relevant studies range from the fundamental properties of electronic materials \citep{Hasan2010} to the engineering of optical lattices in cold atom systems~\cite{Cooper2019}, and from  topological photonics~\cite{Ozawa2019} to applications in phononic and acoustic systems~\cite{Susstrunk2016, Ma2019}. A central role in such works has been played by the so-called ``bulk-boundary correspondence''~\cite{Bernevig2013}. This has enabled an understanding (based on infinite/bulk materials) of how finite or semi-infinite systems may behave in the presence of corners, edges, and surfaces~\cite{Kane2005, Wan2011, Benalcazar2017}. 

The present work treads at the nexus of all three above directions. In particular, we aim to examine a dimer system of the FPUT type. Exponentially localized in space, temporally periodic in time solutions in the form of the so-called discrete breathers~\cite{Flach2008, Aubry2006} have been identified in such systems in several earlier works~\cite{Livi_1997,maniadis, JAMES2004124}. Typically, in such problems, a variation of the mass between the elements of the dimer lattice is considered. Here, however, we consider a model in the spirit of numerous works motivated by the so-called  Su-Schrieffer-Heeger (SSH) model~\cite{Su1979}, a recent popular platform for controlling the band gap features and associated edge modes~\cite{Ozawa2019}. More concretely, we explore a dimer in the linear couplings~\cite{MS2021, CXYKT2021} while we preserve the softening nonlinearity of the same ($\beta$-FPUT type) across all the bonds (intersite) of the lattice. It is worth noting that a very similar linear setting, but for an onsite (rather than intersite) nonlinearity, has been very recently explored in Ref.~\cite{hofstrand}. This work aims to provide an analysis (using both continuum methods and direct numerical simulations) of the prototypical nonlinear patterns that can arise in bulk and the edges of the nonlinear lattice model under consideration.

Our presentation is structured as follows. First, Section II introduces the model and briefly discusses its properties in the linear regime. Then, in Section III, leveraging a formal continuum limit, we are led naturally to a nonlinear Dirac equation. We find that the nonlinearity of the derived Dirac model does not match well-established cases, such as the Soler/Gross-Neveu or the (integrable) massive Thirring model~\cite{our_nab,nabil2}. Nevertheless, inspired by related work, such as that of Ref.~\cite{SSLK2019}, we devise a sequence of linear and nonlinear transformations that ultimately allow us to compute the stationary soliton of the continuum approximation. Part of our motivation for developing the relevant reduction stems from the existence of established stability criteria for PDEs of the nonlinear Dirac type~\cite{Berkolaiko_2015} that we intend to leverage to suggest the stability of the identified waveforms. In Section IV, we explore how to adapt the relevant solutions to the context of a semi-infinite (i.e., with one end being bounded) continuum. In Section V,  we analyze the nonlinear solutions in the bulk and edges of the finite lattice and compare them with ones obtained from solving the PDEs. Finally, in Section VI, we summarize our findings and present some exciting directions for future studies. The Appendices complement our presentation with some of the technical details of the system.

\section{Model Setup}  \label{Section1}
\begin{figure}[t!]
	\centering
	\includegraphics[width=\columnwidth]{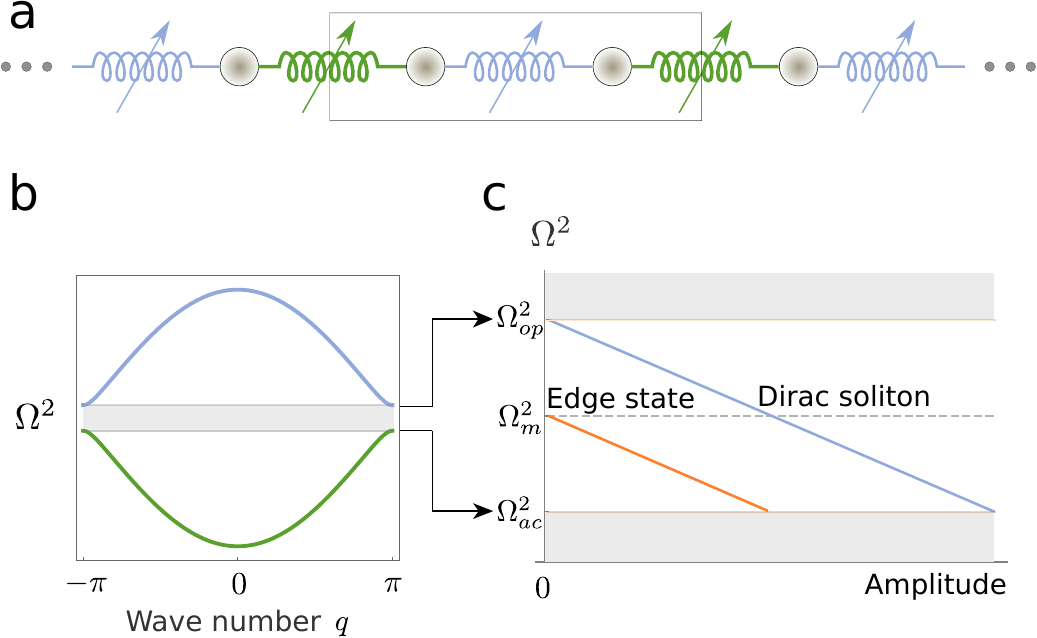} 
	\caption{
 (a) FPUT dimer lattice with equal masses connected with nonlinear springs with two different linear stiffnesses. The box denotes the unit cell. 
 (b) Dispersion diagram of the linear system (i.e., when the nonlinear
coefficient $\Gamma=0$). 
(c) A schematic of the amplitude-dependent edge state and the Dirac soliton residing inside the band gap.}
	\label{fig1} 
\end{figure}

We consider a periodic chain made of two alternating springs with a weak cubic nonlinearity, as shown in Fig.~\ref{fig1}a. The non-dimensional equations of motion for the two particles inside the $n$th unit cell can be written as follows \cite{CT2019}:  
\begin{equation}
\left.
\begin{IEEEeqnarraybox}[
\IEEEeqnarraystrutmode
\IEEEeqnarraystrutsizeadd{2pt}
{2pt}
][c]{rCl}
\ddot{\xi}_{1,n} &=& (1+\gamma) (\xi_{2,n-1}-\xi_{1,n})+\mathrm{\Gamma} (\xi_{2,n-1}-\xi_{1,n})^3  \\ 
&& -\>(1-\gamma) (\xi_{1,n}-\xi_{2,n})-\mathrm{\Gamma} (\xi_{1,n}-\xi_{2,n})^3,   \\
 \ddot{\xi}_{2,n}&=&(1-\gamma) (\xi_{1,n}-\xi_{2,n})+ \mathrm{\Gamma} (\xi_{1,n}-\xi_{2,n})^3  \\ 
&&- \>(1+\gamma) (\xi_{2,n}-\xi_{1,n+1})- \mathrm{\Gamma} (\xi_{2,n}-\xi_{1,n+1})^3.
\end{IEEEeqnarraybox}
\, \right\}
\label{eq:eom_FPU}
\end{equation}
Here, $\xi_{m,n}$ denotes the normalized displacement of the $m$th particle inside the $n$th cell, $1-\gamma$ and $1+\gamma$ represent the linearized stiffness of two springs, and $\mathrm{\Gamma}$ is the nonlinearity parameter. We take the same nonlinearity parameter for all springs to make the analytical treatment simpler.

In the linear limit ($\mathrm{\Gamma} \rightarrow 0 $), the system represents a periodic chain consisting of two alternating springs with stiffness $1-\gamma$ and $1+\gamma$. The dispersion relation has two branches as shown in Fig.~\ref{fig1}b. By assuming  $\gamma > 0$, at the edge of  Brillouin zone (BZ), i.e., at wavenumber $q= \pm \pi$, the acoustic (lower) and optical (upper) cutoff frequencies are respectively given by: 
$$\w_{ac}^2 = 2(1-\gamma), \quad  \w_{op}^2 = 2(1+\gamma),$$ %respectively. 
Thus, the dispersion curve has a band gap (defined in terms of frequency square here) of width  $ \w_{op}^2-\w_{ac}^2$. 
Moreover, the eigenmode corresponding to $\w_{ac}^2$ is given by $$(\xi_{1,n}, \xi_{2,n})=(A, -A) \exp(i  q n)=(-1)^n(A, -A),$$ where $A$ is the amplitude of oscillation. Physically, this means that the two particles inside the unit cell oscillate \textit{out-of-phase}. Similarly, the eigenmode at $\w_{op}^2$ is given by: 
$$(\xi_{1,n}, \xi_{2,n})=(A, A)\exp(i  q n)=(-1)^n(A, A),$$ representing the \textit{in phase} motion of particles inside the unit cell. The prefactor $(-1)^n$ is due to the corresponding wavenumber. We will utilize these characteristics to discover amplitude-dependent solitons and edge states residing in this band gap, as illustrated in Fig.~\ref{fig1}c. Having set up the relevant model, we now turn to the analysis of its prototypical soliton solutions over the infinite lattice.

\section{Infinite continuum}  \label{Section2}
We focus on the weakly nonlinear wave solutions, i.e., $ \mathrm{\Gamma}=\epsilon_1 \Tilde{\mathrm{\Gamma}}$ with $\epsilon_1 \ll 1$, inside the band gap for the wave number $q=\pi$. Moreover, we consider a small band gap, such that $\gamma = \epsilon_2 \Tilde{\gamma}$ with $\epsilon_2 \ll 1$. We further assume that nonlinearity and band gap are of the same order, i.e., $\epsilon_1 = \epsilon_2 = \epsilon$. We then look for slowly-varying solutions around frequency $\w$. The structure of the two eigenmodes at $q=\pi$ suggests that we look for solutions with the following ansatz:
\begin{equation}
\left.
\begin{IEEEeqnarraybox}[
\IEEEeqnarraystrutmode
\IEEEeqnarraystrutsizeadd{2pt}
{6pt}
][c]{rCl}
\xi_{1,n} =\frac{ (-1)^n }{2} \left [u(z,\tau) \exp(i \mathrm{\w} t) + u^*(z,\tau) \exp(-i \mathrm{\w} t) \right], \\
\xi_{2,n}= \frac{ (-1)^n }{2} \left [v(z,\tau) \exp(i \mathrm{\w} t) + v^*(z,\tau) \exp(-i \mathrm{\w} t) \right], 
\end{IEEEeqnarraybox}
\, \right\}
\label{ansatz_FPU}
\end{equation}
where star denotes complex conjugate, and $z = \epsilon n$ and $\tau = \epsilon t $. We substitute the ansatz in Eq.~\eqref{eq:eom_FPU} and proceed formally to a continuum approximation; notice that, to do so, we are partly motivated by its successes in similar problems~\cite{Kivshar1992, Chubykalo1993b} and partly through an {\it a posteriori} comparison with the lattice dynamical results. We thus consider that the functions $u$ and $v$ are approximated by continuous functions of the position (and time), and expanding in the Taylor series, we equate the various orders of $\epsilon$.

At $\order{\epsilon^0}$, we obtain $\w = \sqrt{2}$. 
%This corresponds to the dispersion curve at $q=\pi$ for a monomer chain with all springs having the same stiffness. 
In the dimer model presented in Section \ref{Section1}, this frequency corresponds to the midgap frequency $\w_m = \sqrt{(\w_{ac}^2 + \w_{op}^2)/2} = \sqrt{2}$. This makes sense because as the stiffness difference $\gamma$ approaches zero (the leading order dynamics), the band gap in our dimer lattice goes to the limit of the monoatomic lattice with frequency $\w = \sqrt{2}$ at $q=\pi$.

At $\order{\epsilon^1}$, however, we obtain the following system of nonlinear PDEs for the solutions around $\w_m = \sqrt{2}$:  
\begin{equation}
\left.
\begin{IEEEeqnarraybox}[
\IEEEeqnarraystrutmode
\IEEEeqnarraystrutsizeadd{2pt}
{6pt}
][c]{rCl}
- 4 i \sqrt{2} \frac{\partial u(z,\tau)}{\partial \tau} &=& -2  \frac{\partial v(z,\tau)}{\partial z}  + 4 \Tilde{\gamma} v(z,\tau) \\
&&+\>  3 \mathrm{\Tilde{\Gamma}} \left [ \abs{u(z,\tau)}^2 + 2 \abs{v(z,\tau)}^2 \right ]u(z,\tau) \\
&&+\> 3 \mathrm{\Tilde{\Gamma}} {v(z,\tau)}^2 u^*(z,\tau),  \\
- 4 i \sqrt{2} \frac{\partial v(z,\tau)}{\partial \tau} &=& 2 \frac{\partial u(z,\tau)}{\partial z}  + 4 \Tilde{\gamma} u(z,\tau) \\
&&+\>  3 \mathrm{\Tilde{\Gamma}} \left [2 \abs{u(z,\tau)}^2 + \abs{v(z,\tau)}^2 \right ]v(z,\tau) \\
&&+\> 3 \mathrm{\Tilde{\Gamma}} {u(z,\tau)}^2 v^*(z,\tau). 
\end{IEEEeqnarraybox}
\, \right\}
\label{eq:pde_FPU}
\end{equation}
In the linear limit, $\mathrm{\Tilde{\Gamma}} \rightarrow 0$, these PDEs closely follow the dispersion curve obtained for the discrete lattice in Eq.~\eqref{eq:eom_FPU} (see Appendix \ref{AppendixA} for comparison). Interestingly, we can further simplify these PDEs using a suitable rotation (together with a simple rescaling) and obtain a bi-spinor nonlinear Dirac (NLD) equation in the form:
\begin{equation}
\left.
\begin{IEEEeqnarraybox}[
\IEEEeqnarraystrutmode
\IEEEeqnarraystrutsizeadd{2pt}
{6pt}
][c]{rCl}
 i \frac{\partial {\psi_1}}{\partial \tau}  &=& -  \frac{\partial {\psi_2}}{\partial s}  - \tilde{M}   {\psi_1} +  {|{\psi_1}|}^2 {\psi_1},  \\
 i \frac{\partial {\psi_2}}{\partial \tau}  &=&   \frac{\partial {\psi_1}}{\partial s}  + \tilde{M}    {\psi_2} +  {|{\psi_2}|}^2 {\psi_2}.
\end{IEEEeqnarraybox}
\, \right\}
\label{eq:NLD_FPU}
\end{equation}
Here,  
$\psi_1 = \tilde{G} (u+v)$, $\psi_2 =  \tilde{G}  (u-v)$, 
$\tilde{G}  = \sqrt{-3 \mathrm{\Tilde{\Gamma}}/(4 \sqrt{2}})$,
$s = 2\sqrt{2} z$, 
$\tilde{M} =\Tilde{\gamma}/\sqrt{2}$.
We considered $\mathrm{\Tilde{\Gamma}}<0$ for the softening nonlinearity of interest herein. Also, the accent ($~\Tilde{}~$) indicates the normalized system parameters of $\order{1}$. It is worth pointing out that these equations are invariant under the transformation $s \rightarrow -s$, $\psi_1 \rightarrow \psi_1$, and $\psi_2 \rightarrow -\psi_2$. Therefore, if the initial conditions of the PDEs $\psi_1(s)$ and $\psi_2(s)$ are \textit{even} and \textit{odd} functions of $s$, respectively, the solution would preserve this symmetry for all times. It is relevant to point out here that the
presence of nonlinearities involving self- and cross-phase
modulation (in the language of nonlinear optics) precludes
the existence of Lorentz invariance in the model derived herein,
contrary, e.g., to what is the case in the setting discussed
in~\cite{ALEXEEVA2019198}.%
Notice that the same procedure can be used to obtain the NLD equations for the KG lattice with the onsite nonlinearity. The latter model has been explored in considerable detail in the very recent work~\cite{hofstrand}.

It is worth mentioning some key differences between this system and the dimer models, in which the envelope dynamics inside the band gap is governed by the nonlinear Schrodinger (NLS) equation~\cite{Huang1998}.  In fact, the latter is the case with large bandgap, $\gamma \approx \order{1}$ and weaker nonlinearity $\mathrm{\Gamma} \approx \order{\epsilon^2}$, where the slow scales $z = \epsilon n$ and $\tau = \epsilon^2 t$ govern the envelope solutions. However, it was shown by Hu et al.~\cite{Huang2000} that a small band gap, as is the case here, could lead to new kinds of gap solutions governed by coupled-mode equations~\cite{Mills1987}.

Next, following Ref.~\cite{SSLK2019}, we further employ the transformation:
\begin{equation}
{\psi_1}  \rightarrow  \left(\psi_a + \psi_b  \right)  
e^{i \tilde{\omega} \tau}, \quad 
{\psi_2}  \rightarrow  -i (\psi_a - \psi_b) e^{i \tilde{\omega} \tau},
 \nonumber
\end{equation}
and obtain
\begin{equation}
	\left.
	\begin{IEEEeqnarraybox}[
		\IEEEeqnarraystrutmode
		\IEEEeqnarraystrutsizeadd{2pt}
		{6pt}
		][c]{rCl}
		\frac{\partial \psi_a}{\partial \tau}  &=&  \frac{\partial \psi_a}{\partial s}  - i  \tilde{\omega} \psi_a +    i \tilde{M}  \psi_b \\
		&& -\> i(  { |\psi_a| }^2 + { |\psi_b| }^2) \psi_a   
		- i (\psi_a \psi_b^* + \psi_a^* \psi_b) \psi_b, \\
		\frac{\partial \psi_b}{\partial \tau}  &=& -  \frac{\partial \psi_b}{\partial s}  + i \tilde{M}  \psi_a-    i \tilde{\omega} \psi_b \\
		&& -\> i(  { |\psi_a| }^2 + { |\psi_b| }^2) \psi_b   
		- i (\psi_a \psi_b^* + \psi_a^* \psi_b) \psi_a,
	\end{IEEEeqnarraybox}
	\, \right\}
	\label{eq:pdeNorm_FPU}
\end{equation}
where $\tilde{\omega}$ can be interpreted as the frequency offset from the midgap frequency $\w_m$ 
%in slow time scale $\tau$. $
% However, the frequency offset is $\omega  = \epsilon \tilde{\omega}$ in the actual time $scale $t$.$
The bandgap region corresponds to $\tilde{\omega} \in [-\tilde{M}, \tilde{M}]$. 
% or ${\omega} \in [-{M}, {M}]$, where $M=\epsilon \tilde{M}$.$
It is relevant to point out here that the resulting class of models of Eq.~(\ref{eq:pdeNorm_FPU}) is strongly reminiscent of the one describing the propagation of slow Bragg solitons in nonlinear refractive periodic media; these models were widely studied over 30 years ago in pioneering studies such as those of~\cite{ACEVES198937, PhysRevLett.62.1746}. It will be techniques for identification of the solitons in such systems that we will leverage to obtain exact solutions for the stationary waveforms in what follows.

% \textit{Important point:
% In our previous version, the linear limit was defined as $\mathrm{\Gamma} \rightarrow 0$. This led to a dimer configuration. However, as per the new notations, the linear limit will be defined by $\Tilde{\mathrm{\Gamma}} \rightarrow 0$, which leads to a dimer configuration again. This is different from $\epsilon \rightarrow 0$, which leads to a monomer.}

%{\bf GT Things to do and check: What about other lattices, DNLS and KG nonlinear lattices? 
%KG lattice can be reduced to DNLS using the same assumptions used in this paper - %rotated wave approximation, weakly nonlinear etc. So, I guess that KG lattices and %DNLS can be reduced to the same ND equation, the one of Smirnova. What about the %FPUT lattices? 
%Using the u-v and u+v transformation the nonlinearity of the reduced ND is also of %Kerr. Do we have the same ND equation?}

\subsection{Stationary solutions}

We now seek stationary solutions for $\tilde{\omega}$, which corresponds to a frequency ${\mathrm{\Omega} = \mathrm{\Omega_m} + \epsilon \tilde{\omega}}$ in the dispersion diagram. Such stationary solutions do not depend on time and, hence, Eq.~\eqref{eq:pdeNorm_FPU} is reduced to the form: 
\begin{equation}
\left.
\begin{IEEEeqnarraybox}[
\IEEEeqnarraystrutmode
\IEEEeqnarraystrutsizeadd{2pt}
{6pt}
][c]{rCl}
\frac{d \psi_a}{d s} &=&  i \Bigl[  \tilde{\omega} \psi_a -  \Tilde{M}  \psi_b + ( | \psi_a|^2 + | \psi_b|^2)  \psi_a     \\
&& + (\psi_a \psi_b^* + \psi_a^* \psi_b) \psi_b \Bigr],  \\
\frac{d \psi_b}{d s} &=&  i  \Bigl[  \tilde{M}  \psi_a  -   \tilde{\omega} \psi_b  - ( | \psi_a|^2 + | \psi_b|^2)  \psi_b  \\
&& - (\psi_a \psi_b^* + \psi_a^* \psi_b) \psi_a \Bigr].
\end{IEEEeqnarraybox}
\, \right\}
\label{eq:odeNorm_FPU}
\end{equation}
Next, we use a polar decomposition into amplitude and phase variables, namely:
\begin{eqnarray}
\psi_a(s) &=& \phi_a (s) \exp \left[i (\theta_0(s)+ \theta(s))/2 \right ], \nonumber \\ 
\psi_b (s) &=& \phi_b(s) \exp \left [i (\theta_0(s) -  \theta(s))/2 \right], \nonumber
\end{eqnarray}
%
%the complex quantities  $\psi_a(s) \equiv \phi_a (s) \exp \left[i (\theta_0(s)+ \theta(s))/2 \right ] $ and $\psi_b (s) \equiv \phi_b(s) \exp \left [i (\theta_0(s) -  \theta(s))/2 \right]$ with $\phi_a (s), \phi_b (s) \geq 0$, i.e., using
%an effective polar decomposition into amplitude and phase variables. We can then 
%
with $\phi_a (s),~\phi_b (s) \geq 0$, and 
arrive at the following four coupled ODEs for $\phi_a(s)$, $\phi_b(s)$, $\theta_0(s)$, and $\theta(s)$:
\begin{eqnarray}
\frac{d \phi_a}{d s} &=&     {\phi_b \sin{ \theta} \left (2 \phi_a \phi_b \cos{\theta}  - \tilde{M}  \right) }, 
\label{a}  \\
\frac{d \phi_b}{d s} &=&   {\phi_a \sin{ \theta} \left (2 \phi_a \phi_b \cos{\theta}  - \tilde{M}  \right) }, 
\label{b} \\ 
\frac{d \theta}{d s} &=&  2 \tilde{\omega} 
  + \left ( \frac{\phi_a}{\phi_b } + \frac{\phi_b}{\phi_a } \right) \\ && \times \left [\cos{\theta}(2 \phi_a \phi_b \cos{\theta} - \tilde{M} ) + 2 \phi_a \phi_b  \right],
\nonumber \\
\label{c} \\
\frac{d \theta_0}{d s} &=&   \left ( \frac{\phi_a}{\phi_b } - \frac{\phi_b}{\phi_a } \right) \left [\cos{\theta} (\tilde{M}  - 2 \phi_a \phi_b \cos{\theta})  \right]. 
\label{d} 
\end{eqnarray}
By dividing Eq.~\eqref{a} by Eq.~\eqref{b}, we get
\begin{eqnarray}
\frac{d \phi_a}{d \phi_b} = \frac{\phi_b}{\phi_a},
\end{eqnarray} 
which, upon integration, yields:
%we can integrate easily and obtain 
\begin{eqnarray}
{\phi_a^2 - \phi_b^2  = c }. \label{eq:integrationConst}
\end{eqnarray}
Here, the integration constant 
%of integral 
$c$ can be found from  
%calculated by imposing the suitable 
the boundary conditions.

\subsection{Dirac soliton}

In this study, we are interested in localized soliton solutions of this continuum approximation, so that we could translate them into approximate solutions (or initial guesses in the context of our numerical computations) of the discrete system breather waveforms. Therefore, we impose the boundary conditions $\phi_a(s) \rightarrow 0$ and $\phi_b(s) \rightarrow 0$ as $s \rightarrow \infty$. This translates into vanishing $u$, $v$, and $\xi$ at infinity, and $c=0$ from Eq.~\eqref{eq:integrationConst}. Assuming $\phi_a(s)$ and $\phi_b(s)$ to be non-negative, we thus have $\phi_a(s) = \phi_b(s)$ for $s \in (-\infty, \infty)$. A solution decaying to zero \textit{also} as $s \rightarrow -\infty$, e.g., a soliton, naturally satisfies this condition.
%However, for a semi-infinite case, any boundary satisfying $\phi_a^2 = \phi_b^2$, with both \textit{not} necessarily vanishing, will also be a solution and signify a nonlinear \textit{edge} state.
%
%We further define 
Further, since $c=0$, we define 
$\rho \equiv \phi_a^2 = \phi_b^2$ 
%(i.e., assuming that $c^2=0$ above) and simplify 
and cast Eqs.~\eqref{a}--\eqref{d} into the form:
%to get 
%
\begin{eqnarray}
\frac{d \rho}{d s} &=&   2 \rho \sin{\theta} \left(   {2 \rho \cos{\theta} }- {\tilde{M} } \right) , \label{4}  \\ 
\frac{d \theta}{d s} &=&  2 \cos{\theta}  \left( {2 \rho \cos{\theta} }- {\tilde{M} } \right) + {4 \rho} + {2 \tilde{\omega}}\label{5},  \\
\frac{d \theta_0}{d s} &=& 0\label{6}.
\end{eqnarray}
%
%We first solve 
Equation~\eqref{6} can readily be integrated, leading to $\theta_0(s)=c_1$. 
This implies that 
\begin{eqnarray}
\psi_a(s) &=& \sqrt{\rho(s)} \exp (i c_1/2) \exp \left(i \theta(s)/2 \right ), \nonumber \\
\psi_b(s) &=& \sqrt{\rho(s)} \exp (i c_1/2) \exp \left(-i \theta(s)/2 \right ), \nonumber
\end{eqnarray}
and, thus, in the NLD picture, we have:
\begin{eqnarray}
\psi_1(s, \tau) &=& 2 \sqrt{\rho(s)} \cos(\theta(s)/2) \exp \left [ i (\tilde{\omega} \tau + c_1/2) \right ], \nonumber \\
\psi_2(s, \tau) &=& 2 \sqrt{\rho(s)} \sin(\theta(s)/2) \exp \left [ i (\tilde{\omega} \tau + c_1/2) \right ]. \nonumber
\end{eqnarray}
%	
%It is easy to see that $c_1$ results in a \textit{temporal} phase of $u$ and $v$, and thus of displacement $\xi$ in Eq.~\eqref{ansatz_FPU}. 
In what follows, we choose $c_1=0$ for simplicity,
leveraging the (overall) gauge invariance of the equations.

\begin{figure}[t!]
	\centering
	\includegraphics[width=\columnwidth]{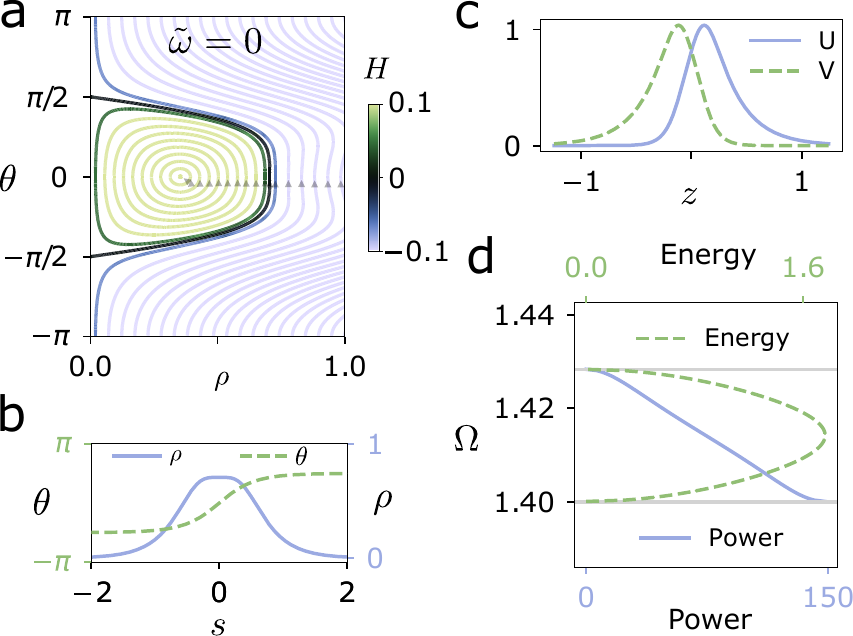} 
	\caption{Infinite continuum NLD findings. (a) Phase portrait for Eqs.~\eqref{4} and \eqref{5} at $\tilde{\omega} = 0$. The colormap denotes the value of the Hamiltonian that corresponds to each trajectory. The heteroclinic orbit, that corresponds to $H=0$, is shown with the black line. (b) Dirac soliton profile ($\rho(s), \theta(s)$) that corresponds to the heteroclinic orbit. (c) Dirac soliton profile ($U(z), V(z)$) (d) Energy and power, the conserved quantities of the NLD equation, obtained for the soliton solution inside the band gap.}
	\label{fig2} 
\end{figure}

The remaining two equations [Eqs.~\eqref{4} and \eqref{5}] in $\rho$ and $\theta$ are decoupled from the third. Therefore, to get better insight, we plot a 2D phase portrait for these equations in Fig.~\ref{fig2}a at a prototypical frequency ($\tilde{\omega}=0$ representing the middle of the gap). 
We choose $\gamma=\mathrm{\Gamma}=0.02$ for all the studies hereafter, which translates to $\epsilon = 0.01$ and $\tilde{\gamma}=\tilde{\mathrm{\Gamma}}=2$.
The phase portrait has a number of fixed points, including the two saddles at $(\rho, \theta) = \left ( 0, \pm \pi/2  \right)$. 
Furthermore, to analytically track the solutions, Eqs.~\eqref{4} and \eqref{5} can be seen as a dynamical system with the following Hamiltonian
\begin{eqnarray}
H(\rho,\theta) &=& - {2  \tilde{\omega} \rho} - {2 \rho^2} + {2 \tilde{M}  \rho  \cos{\theta}}  - {2 \rho^2 \cos^2{\theta}}. \label{eq:hamiltonian}
\end{eqnarray}
%%%%%%%%%%%%%%%%%%%%%%%%%%%%%%%%%%%%%%%
Since we are looking for %an amplitude 
a localized solution, i.e., $\rho\rightarrow 0$ for $s\rightarrow \pm \infty$, this solution is represented by the heteroclinic orbit, namely the trajectory that connects the two saddle points for which $\rho=0$. The corresponding value of the Hamiltonian for this trajectory is zero, $H=0$.
%\GT{I would prefer to plot the Dirac soliton in $\phi$ representation, with both the amplitude and its phase instead of $u,v$ in Fig2(b) to reveal this topological feature.}, 
We thus obtain
\begin{eqnarray}
\rho = \frac{ \tilde{M}   \cos{\theta} - \tilde{\omega}  } { 1+\cos^2{\theta} },
\label{7}
\end{eqnarray}

\noindent which we substitute in Eq.~\eqref{5} to get 
\begin{eqnarray}
\frac{d \theta}{d s} = 2 \left({ \tilde{M}   \cos{\theta} - \tilde{\omega}} \right).
\end{eqnarray}
By integrating, we obtain: 
%get 
\begin{eqnarray}
\theta(s) =  2  \tan^{-1}  \Biggl[  &&  \frac{ \tilde{M}   - \tilde{\omega}}{\sqrt{\tilde{M}^2 - \tilde{\omega}^2}} \tanh \left[ {\sqrt{ \tilde{M}^2 - \tilde{\omega}^2} }  (s - s_0)   \right] \Biggr].
\nonumber \\
\label{eq:theta_galilean}
\end{eqnarray}
Here $s_0$ is the constant of integration. For the soliton solutions in the domain $s \in (-\infty, \infty)$,  we can choose any $s_0$ because those represent shifted members of the 
family of solitons (rendered possible due to the translational
invariance of the underlying model). For convenience, we choose a soliton centered at $s=0$, therefore
\begin{eqnarray}
\theta(s) =  2  \tan^{-1}  \Biggl[  &&  \frac{ \tilde{M}   - \tilde{\omega}}{\sqrt{\tilde{M}^2 - \tilde{\omega}^2}} \tanh \left[ {\sqrt{ \tilde{M}^2 - \tilde{\omega}^2} }  s   \right] \Biggr].
\nonumber \\
\label{eq:stationarysoliton}
\end{eqnarray}
%We use Eqs.~\eqref{6} and \eqref{7} to get
%\begin{eqnarray}
%\rho(s) &=& \left( \frac{M \cos{\theta}  - \omega  } { 1+  %\cos^2{\theta} } \right).
%\end{eqnarray}
In Fig.~\ref{fig2}b, we show the soliton profile. We note that the amplitude vanishes as $s \rightarrow \pm \infty$. However, the phase reverses from $-\pi/2$ to $\pi/2$. The latter indicates its similarity with a topological soliton.

Finally, $\rho(s)$ and $\theta(s)$ can be used to obtain
\begin{eqnarray}
\psi_a(s) &=& \sqrt{\rho(s)} \exp \left (i \theta(s)/2 \right), \\ 
\psi_b(s) &=& \sqrt{\rho(s)} \exp \left (-i \theta(s)/2 \right), \\
\psi_1(s, \tau) &=& 2 \sqrt{\rho(s)} \cos(\theta(s)/2) \exp(i \tilde{\omega} \tau), \\
\psi_2(s, \tau) &=& 2 \sqrt{\rho(s)} \sin(\theta(s)/2) \exp(i \tilde{\omega} \tau), \\
u(z, \tau) &=& U(z) \exp(i \tilde{\omega} \tau), \\
v(z, \tau) &=& V(z) \exp(i \tilde{\omega} \tau), 
\end{eqnarray}
where $s = 2\sqrt{2} z$, and 
\begin{eqnarray}
U(z) &=& \frac{\sqrt{2 \rho(s)}}{\tilde{G} }
\sin \left( \frac{\pi}{4} + \frac{\theta(s)}{2}  \right), \quad
%$, and $
\nonumber \\
V(z)&=&\frac{\sqrt{2 \rho(s)}}{\tilde{G} }
\sin \left( \frac{\pi}{4} - \frac{\theta(s)}{2}  \right). 
\nonumber
\end{eqnarray}
In Fig.~\ref{fig2}c, we show the soliton profile at $\tilde{\omega}=0$ in terms of $U(z)$ and $V(z)$.

It is relevant to point out here (both for analytical and for numerical
purposes) that it is possible to identify other trajectories of the dynamical system as well. In particular, a positive finite value of $H$ leads to a quadratic equation for $\rho$ that can be solved explicitly in terms of
$\theta$ and back-substituted into the ODE for $d\theta/ds$ in order
to retrieve the corresponding periodic orbits from the integration of
the ODE for $\theta=\theta(s)$.

\subsection{Conserved quantities}
We now discuss the frequency dependency of the conserved quantities of the NLD equations shown in Eq.~\eqref{eq:NLD_FPU}.
First, the power of the NLD equations is given as
\begin{eqnarray}
P(\tilde{\omega}) = \left (\frac{1}{ 4 {G}^2 \sqrt{2} } \right )\int_{-\infty}^{\infty} (|{\psi}_1|^2 + |{\psi}_2|^2) \text{d}s,
\label{eq:power}
\end{eqnarray}
where the factor in front of the integration, which is independent of the frequency, is introduced to scale the expression and compare it with the total lattice energy of the stationary solutions of Eq.~\eqref{eq:eom_FPU} (see Appendix \ref{AppendixB} for details). For the stationary soliton [Eq.~\eqref{eq:stationarysoliton}], the power can thus be deduced to be:
\begin{eqnarray}
P(\tilde{\omega}) = \left(\frac{1}{2 {G}^2} \right) \Bigg[ &&\tan^{-1} \left( {1+\sqrt{\frac{2(\tilde{M}-\tilde{\omega})}{\tilde{M}+ \tilde{\omega}} }} \right) \nonumber \\
- &&\tan^{-1} \left( {1-\sqrt{\frac{2(\tilde{M}-\tilde{\omega})}{\tilde{M}+\tilde{\omega}} }} \right) \Bigg],
\end{eqnarray}
where $G^2 = \epsilon {\tilde{G}}^2$.
In Fig.~\ref{fig2}d, we plot the power, which increases monotonically with the decrease in frequency. 
In line with the work of~\cite{Berkolaiko_2015}, the Vakhitov-Kolokolov criterion about the 
sign of the derivative of $P$ with $\tilde{\omega}$ here 
suggests the stability of the solitons.

Similarly, we write another key conserved quantity, the energy $E$, for  Eq.~\eqref{eq:NLD_FPU} as
\begin{eqnarray}
E =  \int_{-\infty}^{\infty}  \bigg [  {\psi}_1^* \frac{\partial {\psi_2}}{\partial s} &-& {\psi}_2^* \frac{\partial {\psi_1}}{\partial s}  + \frac{\tilde{M}}{2}(|{\psi}_1|^2 - |{\psi}_2|^2) \nonumber \\
&-& \frac{1}{2}(|{\psi}_1|^4 + |{\psi}_2|^4) \bigg] \text{d}s.
\end{eqnarray}
For the stationary soliton [Eq.~\eqref{eq:stationarysoliton}], the energy reduces to
\begin{eqnarray}
E(\tilde{\omega}) =  \sqrt{2} \tilde{M} \tanh^{-1} \left (\sqrt{\frac{1}{2} \left(1 - \frac{\tilde{\omega}^2}{\tilde{M}^2} \right)} \right).
\end{eqnarray}
In Fig.~\ref{fig2}d, we observe that, as opposed to the power, the energy changes non-monotonically as a function of frequency. Yet, it is important to 
point out that the energy maintains a definite sign and does not
have a zero crossing. This is also in line with the absence of instability
according to the second criterion
of the work of~\cite{Berkolaiko_2015} in the context of Dirac
equations. This criterion associates the zero crossings of the energy
with a change of stability. 

In summary, both stability criteria associated with nonlinear Dirac PDEs suggest the absence of instabilities for the localized waveforms examined herein. While, given the reduced nature of the 
%effective 
NLD equation, the stability findings from these criteria are merely suggestive of the  absence of a point spectrum (i.e., isolated linearization eigenvalue pair) instability for the breathing waveforms identified herein, we will see below that our numerical computations corroborate such findings. As an aside, we note that the %effective 
translational invariance of the NLD PDE is tantamount to the conservation of linear momentum. Yet, since this latter conservation law is not directly related to the stability criteria of~\cite{Berkolaiko_2015}, we do not examine the latter in detail herein.

%{\bf What about linear momentum? The translational invariance
%of the model suggests that linear momentum would also be a conserved 
%quantity.From Eq.~(8) of~\cite{Berkolaiko_2015}, it seems that the
%linear momentum is exactly like the one of NLS and should be
%conserved. Do we discuss it/do we compute it? Or we content
%ourselves with saying that the Berkolaiko criteria involve
%only $dP/d\omega$ and the sign of $E$ and we leave it at that?}
%\GT{I would say to follow the second proposition - comment and leave it at that - but lets discuss it}

%\GT{All this discussion about calculating the energy and power of the nonlinear Dirac was initiated in order to make the connection with the existed bibliography about stability of nonlinear Dirac equations. To check again these papers and make some comments about the stability - in a level of pdes- of the obtained Dirac solitons}

%\GT{Another tricky point is why we went to Dirac formalism and did not stay with the pdes of u and v? There also a phase plane analysis can give soliton solutions etc. The use of Energy and power of nonlinear Dirac could be an answer. Daria uses also some "spin" textures/concepts also, if I remember.} \RC{I think the issue is that $u$ and $v$ could be complex functions in space. Going to $\phi$ and $\theta$ representation leads to pdes for real functions.}

\section{Semi-infinite continuum}   \label{Section3}
Up to now our approach has been general in terms of obtaining localized solutions in the continuum approximation of the infinite lattice limit. We now seek decaying (edge) solutions for a \textit{semi-infinite} domain $s \in [0, \infty)$. Recall that Eq.~\eqref{eq:theta_galilean} was derived by imposing vanishing amplitudes only at $s \rightarrow \pm \infty$. Therefore, we can construct nonlinear edge solutions (finite at one edge and decaying as one moves farther away) with the same expression as given in Eq.~\eqref{eq:theta_galilean} as long as $\theta(0)$ satisfies the given boundary condition at the edge. 

Though any $\theta$ can be chosen at the boundary to obtain the corresponding edge solution, we are interested in some special cases, e.g., $\theta(0) = - \pi/2$ ($U(0)=0$), $\theta(0) = 0$ ($U(0)=V(0)$), and $\theta(0) = \pi/2$ ($V(0)=0$). The physical meaning of such boundary conditions will be evident in the next section when we deal with a finite discrete chain. However, before moving further, we can reach some important conclusions. 

We know that $\theta(s)$ varies spatially from ${\theta(s)=-\cos^{-1}(\tilde{\omega}/\tilde{M})}$ to $\theta(s) = \cos^{-1}(\tilde{\omega}/\tilde{M})$ in an infinite continuum [note the saddle points for Eqs.~\eqref{4} and \eqref{5}]. Also, $\tilde{\omega} \in [-\tilde{M}, \tilde{M}]$ inside the band gap. Therefore, $\theta(s)$ in Eq.~\eqref{eq:theta_galilean} obtained for an infinite chain will satisfy the boundary conditions $\theta(0) = - \pi/2$ ($U(0)=0$) and $\theta(0) = \pi/2$ ($V(0)=0$) \textit{only} for $\tilde{\omega} \leq 0$, i.e., below the mid gap frequency $\w_m$. However, the boundary condition $\theta(0) = 0$ ($U(0)=V(0)$) will be satisfied for all $\tilde{\omega}$ inside the band gap.

\begin{figure}[t!]
	\centering
	\includegraphics[width=\columnwidth]{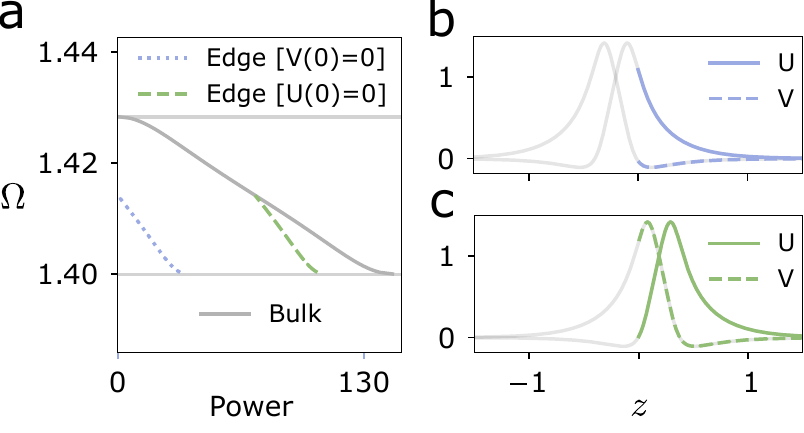} 
	\caption{Semi-infinite continuum NLD results. (a) Power of nonlinear edge states and Dirac solitons inside the band gap. 
 (b) Edge state (nonlinearity-modified) for the boundary with $\theta(0) = \pi/2$ ($V(0)=0$) at $\tilde{\omega} = -\tilde{M}/2$. The profile is a truncation of a shifted Dirac soliton (grey). 
 (c) Edge state (nonlinearity-induced) for the boundary with $\theta(0) = -\pi/2$ ($U(0)=0$) at $\tilde{\omega} = -\tilde{M}/2$. The profile is the other truncated part of a shifted Dirac soliton (grey).}
	\label{fig3} 
\end{figure}

\subsection{Boundary with $\theta(0) = \pi/2$ ($V(0)=0$)}
For $\theta(0)=\pi/2$, Eq.~\eqref{eq:theta_galilean} yields
\begin{eqnarray}
	s_0 =  - \frac{1}{\sqrt{\tilde{M}^2 - \tilde{\omega}^2}} \tanh^{-1} \left (\sqrt{ \frac{\tilde{M} + \tilde{\omega}}{\tilde{M} - \tilde{\omega}} }\right)\  \forall \  \tilde{\omega} \leq 0.
	\label{eq:fixed2}
\end{eqnarray}
We observe that a nonzero $s_0$ simply means that we have a Dirac soliton which is moved to the $-s$ axis by a distance of $s_0$ ($=2 \sqrt{2} z_0$). We, therefore, get the edge solution for the domain $s \in [0, \infty)$. We calculate the corresponding power from Eq.~\eqref{eq:power} by changing the integration limits to account for the finite boundary. In Fig.~\ref{fig3}a, we show the power as a function of frequency by a blue dashed line. Such edge solutions exist only for $\tilde{\omega} \leq 0$. In the linear limit with vanishing power, these solutions tend to $\tilde{\omega} = 0$, the mid-gap frequency. We show the Dirac soliton in grey in Fig.~\ref{fig3}b for $\tilde{\omega} = - \tilde{M}/2$. Since the profile intersects $V=0$ at a finite $z$ for $\tilde{\omega} < 0$, nonlinear edge states can be considered a part of Dirac solitons, as shown in blue.  

Interestingly, in the linear limit, the intersection of the soliton profile with $V=0$ occurs at $z(\text{or} \ s)  \rightarrow \infty$ for $\tilde{\omega}=0$. From Eq.~\eqref{eq:stationarysoliton}, we know for $\tilde{\omega}=0$, we get $\theta \rightarrow \pi/4$ as $s \rightarrow \infty$. Therefore, from Eq.~\eqref{4}, we deduce: $\rho  \propto \exp{-2 \tilde{M} s}$, or in NLD setting, $[\psi_1, \psi_2]  \propto \exp{- \tilde{M} s}$. This is the well-known Jackiw-Rebbi solution~\cite{Jackiw1976, Cooper2019} in the linear Dirac framework; however, it is an \textit{edge} solution -- different from the standard \textit{interface} solution between two media with different Dirac masses. We conclude that when nonlinearity is involved, this edge state is modified, and a family of solutions is generated, as shown in blue in Fig.~\ref{fig3}a. We call them
\textit{nonlinearity-modified} edge states.  

\subsection{Boundary with $\theta(0) = - \pi/2$ ($U(0)=0$)}
Similarly, if we have $\theta(0)= - \pi/2$, Eq.~\eqref{eq:theta_galilean} yields
\begin{eqnarray}
s_0 =  \frac{1}{\sqrt{\tilde{M}^2 - \tilde{\omega}^2}} \tanh^{-1} \left (\sqrt{ \frac{\tilde{M} + \tilde{\omega}}{\tilde{M} - \tilde{\omega}} }\right)\  \forall \  \tilde{\omega} \leq 0,
\label{eq:fixed1}
\end{eqnarray}
This is equivalent to a Dirac soliton having moved to the $+s$ axis by a distance
of $s_0$. 
In Fig.~\ref{fig3}a, we show the power as a function of frequency by a 
green dashed line. As discussed earlier, such edge solutions exist only for $\tilde{\omega} \leq 0$. At $\tilde{\omega} = 0$, these bifurcate from the Dirac soliton that lies in the bulk.
Note that the bifurcation point corresponds to $s_0 \rightarrow \infty$. 
This means that the edge solution tends to the whole spatial profile of the Dirac soliton, and therefore, their powers tend to  have the same value at the bifurcation point.

Contrary to the edge states discussed in the previous subsection, the edge states, in this case, do not have any linear counterparts for vanishing power. Therefore, these spontaneously arise due to nonlinearity for $\tilde{\omega} \leq 0$. Figure~\ref{fig3}c highlights the profile of such \textit{nonlinearity-induced} edge states. These are reminiscent of nonlinear edge states found in diatomic lattices with two different masses~\cite{Kivshar1998}.

%{\bf From a structural perspective it is not fully clear how these solutions are different than the previous ones. I think there needs to be a more comprehensive explanation of their difference. There are useful hints but I think something more definitive is needed (after reading the next subsection, the more definitive explanation is there and needs to be shifted here, IMO).}

Lastly, we calculate the edge states for the boundary with $\theta(0) = 0$ ($U(0)=V(0)$), which exists for the entire band gap (Appendix~\ref{AppendixC}).
Again, these states are also \textit{nonlinearity-induced} edge states with no linearized edge state at vanishing powers.

\section{Finite lattice}   \label{Section4}
After analyzing the bulk and edge solutions in an infinite and semi-infinite continuum, we now consider a \textit{finite} discrete lattice and calculate nonlinear solutions inside the band gap. In addition, we investigate the instabilities that cause the localized solution to delocalize in space. 

It is known that the FPUT lattice shown in Fig.~\ref{fig1}a in its \textit{linear} limit corresponds to a finite-frequency SSH chain~\cite{CXYKT2021}. Such a lattice supports topologically-protected edge states in the case of fixed boundary conditions. However, this happens when the boundary is symmetry-preserving, which physically means that it does \textit{not} cut the unit cell. By contrast, a symmetry-breaking boundary, which cuts the unit cell, does not support an edge state~\cite{CT2019}. It is at this point that we recognize the physical interpretation of specific boundary conditions that we chose in the last section. 
%{\bf I think that this explanation belongs to the previous section
%(perhaps even 2 sections ago, but most likely in the previous (sub)section.}
The boundary that supports a topological edge state in the linearized finite lattice resembles $V(0)=0$ in the continuum limit. When nonlinearity is turned on, this edge state is referred as \textit{nonlinearity-modified} edge state. 
Similarly, the boundary that does \textit{not} support a topological edge state in the linearized finite lattice resembles $U(0)=0$ in the continuum limit. However, when nonlinearity is turned on, we witness \textit{nonlinearity-induced} edge states at \textit{finite} power. In this section, we will show such edge states in the finite lattice and how they compare with their continuum counterparts.

\begin{figure}[t!]
	\centering
   \includegraphics[width=\columnwidth]{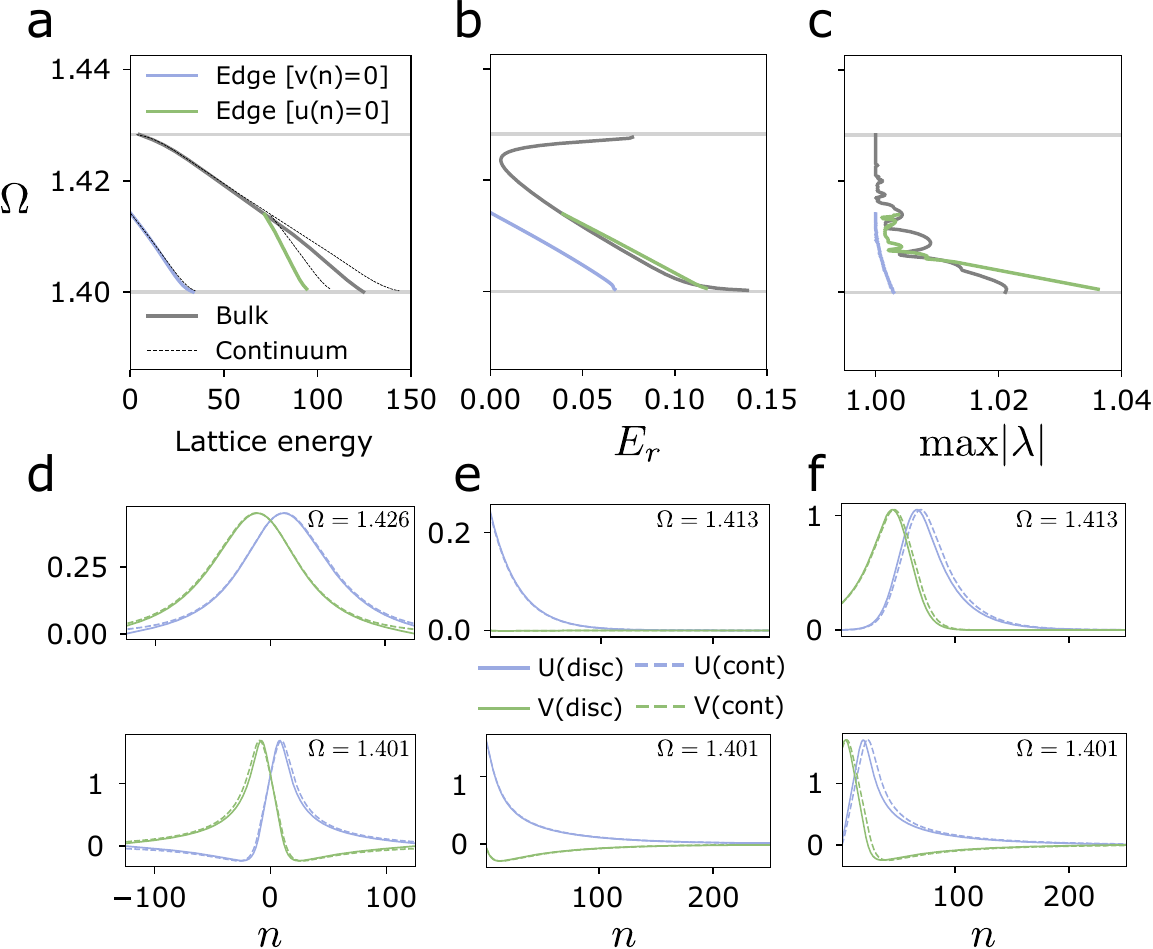} 
	\caption{Finite lattice results. 
 (a) Bifurcation diagram obtained through the nonlinear continuation of the linear edge state (blue) and the first optical state (grey). In green is the nonlinearity-induced edge state. Their lattice energy is compared with the nonlinear states found for the continuum (dashed). 
 (b) Relative energy difference $E_r$ between discrete and continuum solutions plotted inside the bandgap for the bulk and edge solutions.
 (c) Maximum amplitude of the corresponding FMs indicating the extent of instability of discrete nonlinear solutions.
 (d) Comparison of bulk solutions at a frequency near the optical band ($\mathrm{\Omega}=1.426$) and acoustic band ($\mathrm{\Omega}=1.401$). 
 (e)-(f) Comparison of nonlinearity-modified and nonlinearity-induced edge states, respectively, at a frequency near the mid gap ($\mathrm{\Omega}=1.413$) and acoustic band ($\mathrm{\Omega}=1.401$).}
 \label{fig4} 
\end{figure}

\subsection{Bifurcation diagrams}
We take a lattice with 500 particles with fixed ends. The right boundary is kept free when obtaining edge solutions on the left edge. We use Newton's method to find the family of nonlinear periodic solutions for the lattice. By considering the linear edge state and the first state in the optical band as our initial guess, we are able to converge to the nonlinear solution and continue it over frequency.
For the nonlinearity-induced edge states, however, this method does not work since there is no linear limit of such solutions. We tackle this by preparing the initial guess near the acoustic band by truncating the bulk solutions at $U \rightarrow 0$ as discussed in Fig.~\ref{fig3}c. Note that we use symmetry-preserving boundaries in the finite lattice for obtaining the nonlinearity-modified edge states. In contrast, symmetry-breaking boundaries are used for obtaining the bulk and nonlinearity-induced edge states.
In this way, we generate the bifurcation diagram for our discrete lattice as shown in Fig.~\ref{fig4}a. We observe a similar trend as predicted earlier in Fig.~\ref{fig3}a with the existence of \textit{discrete breather} analogs of the Dirac (bulk) solitons, nonlinearity-modified edge states, and nonlinearity-induced edge states. 

To quantify the difference of lattice energy between the continuum model ($E_{\text{cont}}$) and the discrete model ($E_{\text{disc}}$), we define a relative energy parameter 
$$E_r 
= (E_{\text{disc}} - E_{\text{cont}})/E_{\text{cont}}.$$
In Fig.~\ref{fig4}b, we show $E_r$ as a function of frequency. We notice that the $E_r \rightarrow 0$ for the nonlinearity modified edge state in its linear limit. This makes sense because $(\gamma, \mathrm{\Gamma}) \approx O(\epsilon)$, and the zeroth-order dynamics is the linear limit at the midgap frequency $\mathrm{\Omega}_m$, where the edge state lies. $E_r$ for the bulk breather also decreases as it moves closer to the optical band. We observe a sudden rise in $E_r$ very close to the optical band, which could be due to the decrease in localization of the bulk breather and its interaction with finite boundaries. 
%Therefore, for $(\gamma, \mathrm{\Gamma}) \approx O(\epsilon)$, we conclude that the finite lattice, too supports the nonlinear solutions predicted by the \textcolor{red}{continuum} approximation, but an increasing level of energy \textcolor{red}{and for frequencies away from $\mathrm{\Omega}_m$} leads to the deviation of the nonlinear solutions in the finite lattice from the continuum limit. 
In Figs.~\ref{fig4}d, ~\ref{fig4}e, and ~\ref{fig4}f, we show the comparison of discrete and continuum solutions for bulk breather, nonlinearity-modified edge state, and nonlinearity-induced edge state, respectively, at different frequencies inside the bandgap. Overall, we observe an excellent match between discrete and continuum solutions, demonstrating that the finite lattice, too supports the nonlinear solutions predicted by the nonlinear Dirac equations for $(\gamma, \mathrm{\Gamma}) \approx O(\epsilon)$.
%The lattice energy of these discrete solutions matches well with that of continuum solutions at low energy levels. 
We would like to highlight that at the bifurcation point (at the mid gap), the edge state resembles the whole spatial profile of Dirac soliton as discussed in the previous section, therefore their energies tend to be the same for the lattice.
%{\bf I am somewhat surprised that the energy ``matches'' that of the bulk wave. This suggests that the portion of the domain that is ``not present'' does not contribute anything energy-wise (which seems strange). Or am I missing sth?}
%{\bf I think that we should think/explain more about why the midgap solution is the one where this coincidence exists.}

We then perform linear stability analysis of the nonlinear solutions of the finite lattice using  Floquet theory~\cite{Aubry2006}. In Fig.~\ref{fig4}c, we plot the maximum amplitude of the Floquet Multipliers (FMs) corresponding to the nonlinear states inside the band gap. Recall that the values of the FMs that are larger than unity (in absolute value) imply the existence of instabilities. We observe that the Dirac soliton remains linearly stable for higher frequencies, near the optical band. However, it becomes generally unstable with the increase in lattice energy at low frequencies. Such instabilities emerge due to the finite size of the lattice, are associated with 
quartets of FMs with modulus larger than unity, and are expected to vanish for large lattices [see Appendix~\ref{AppendixD} for more details].  This is in line with the fact observed previously (when 
calculating the conserved quantities at the continuum level) that
none of the criteria for the emergence of (in that case, real FM-associated) instabilities of~\cite{Berkolaiko_2015} were met in this context.
%\GT{Hopefully, in the previous section we will say something about the stability of Dirac solitons in the level of PDEs. So in this section, we should highlight that there are instabilities which are due to finite size effects and which disappear in the large lattice limit, making the connection of stability of the Dirac solitons from PDEs and stability of the discrete Dirac solitons.}
Similarly, we also observe that  both types of edge states become unstable with the increase in lattice energy. In particular, nonlinearity-induced edge states are more unstable compared to nonlinearity-modified edge states.
However, in both cases, once again the instabilities (that are progressively
featuring higher growth rates as the acoustic band is approached) are
associated with complex FM quartets, i.e., they are oscillatory in 
nature; see also~\cite{CXYKT2021}.

\subsection{Transient dynamics of discrete Dirac solitons and edge states}
\begin{figure}[t!]
	\centering
   \includegraphics[width=\columnwidth]{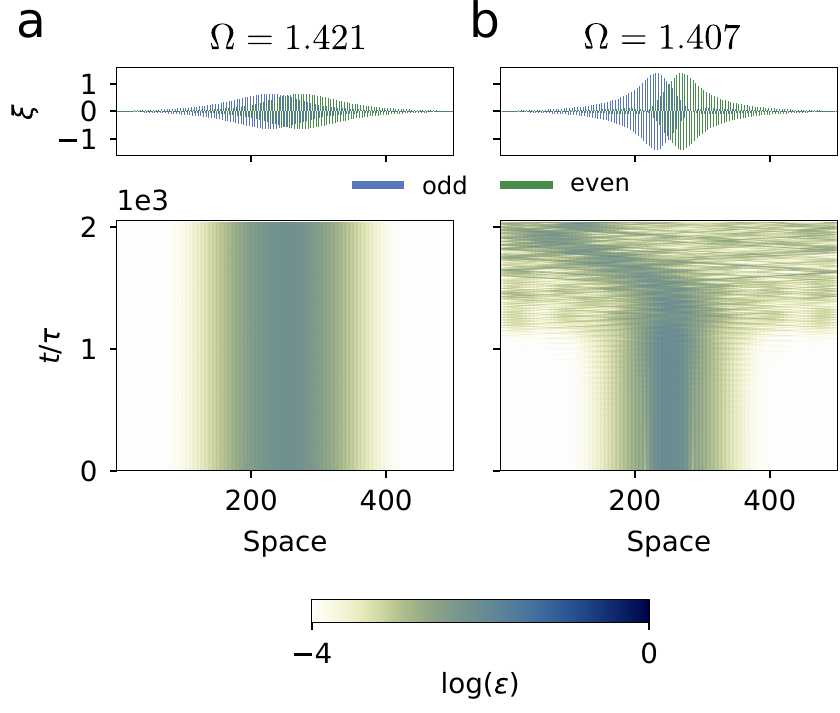} 
	\caption{Transient dynamics of discrete Dirac solitons in the finite lattice. (a) Soliton profile at $\w=1.421$ (even and odd locations are highlighted with different colors) and the transient dynamics when the profile is given as the initial condition to the lattice. Colormap shows the energy density. (b) The same for the soliton at $\w=1.407$, which has a larger FM. The soliton delocalizes due to the presence of instabilities.}
	\label{fig5} 
\end{figure}
We now present the transient dynamics of our discrete analogs
of the continuum Dirac solitons and also of the edge states. In Fig.~\ref{fig5}a, we show a discrete Dirac soliton at $\w=1.421$, which is linearly stable. We apply 1$\%$ noise to its profile and provide the
resulting profile as an initial condition to our finite lattice for a simulation time of $2000 \tau_0$, where $\tau_0$ is the time period of the nonlinear state. We plot the energy density $\varepsilon$ for each mass that includes its kinetic energy and the mean of the potential energy of its left and right neighboring springs, such that
%\begin{widetext}
\begin{eqnarray}
\varepsilon_i
= \frac{\frac{1}{2} \dot{\xi}_i^2 + \frac{1}{2} \left(\text{PE}(\xi_{i-1}, \xi_{i}) + \text{PE}(\xi_{i}, \xi_{i+1}) \right)}{ \sum_{i = 1}^{N} \left(\frac{1}{2} \dot{\xi}_i^2 + \frac{1}{2} \left(\text{PE}(\xi_{i-1}, \xi_{i}) + \text{PE}(\xi_{i}, \xi_{i+1}) \right) \right)}
\end{eqnarray}
%\end{widetext}
where $\text{PE}(\xi_{i}, \xi_{i+1}) = \frac{1}{2} (1 \pm \gamma) (\xi_{i} - \xi_{i+1})^2  +  \frac{1}{4}  \mathrm{\Gamma}(\xi_{i} - \xi_{i+1})^4 $.
We observe that the discrete Dirac soliton remains localized confirming its linear stability. In Fig.~\ref{fig5}b, we show a discrete Dirac soliton at $\w=1.407$, which is linearly unstable through the FM quartets discussed above. Contrary to the previous case, the Dirac soliton starts shedding its energy at around $1200\tau_0$. Interestingly, a localized traveling wave packet is observed as a consequence. 
%We conjecture that these belong to smaller energy and higher frequency soliton solutions, corresponding to the stable regime of Dirac solitons.
Exploring the question of potentially genuine traveling such states is
an interesting question for future work, as we also highlight below
in the Conclusions section.

\begin{figure}[t!]
	\centering
    \includegraphics[width=\columnwidth]{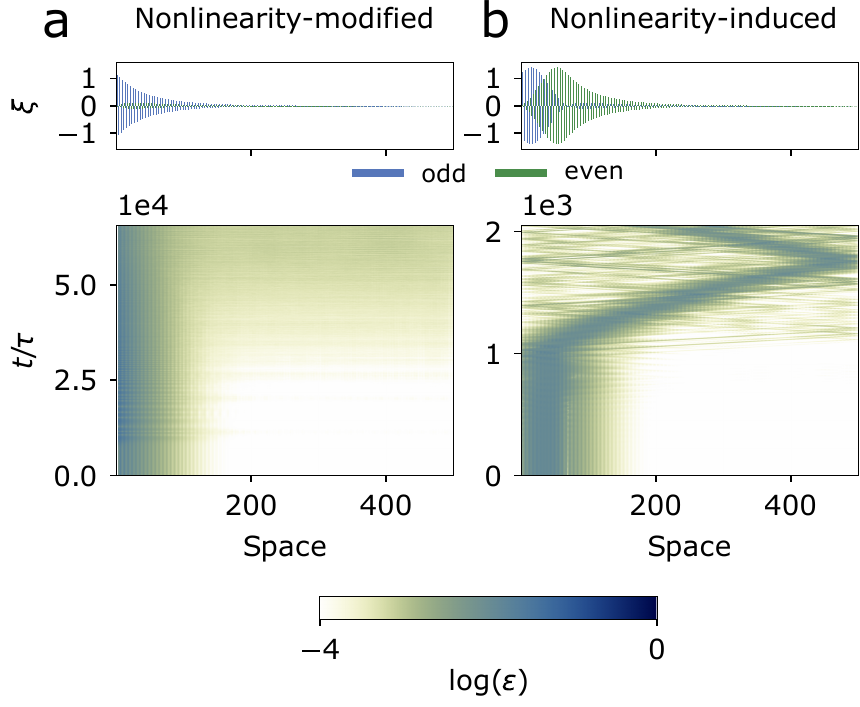} 
	\caption{Transient dynamics of edge states in the finite lattice. (a) Nonlinearity-modified edge state at $\w=1.407$. The transient dynamics shows that the edge state loses its energy to the bulk due to instabilities. (b) Nonlinearity-induced edge state profile and its dynamics at $\w=1.407$. Its transient dynamics shows a stronger delocalization in the form of a robust wave packet moving in the bulk.}
	\label{fig6} 
\end{figure}

Finally, we examine the transient dynamics of nonlinear edge states. In Fig.~\ref{fig6}a, we show a nonlinearity-modified edge state at $\w=1.407$, which is linearly unstable. Recall that a topological edge state exists for this lattice in the linear limit. Transient simulations reveal that the localized mode starts shedding its energy to the bulk gradually while degenerating toward the corresponding linear state. 
%Consequently, we expect that the frequency of the edge state would tend to the frequency of the linearized edge state of the renormalized dispersion for long times~\cite{CXYKT2021, MCTS2022}.
In Fig.~\ref{fig6}b, we show a nonlinearity-induced edge state at the same frequency. Recall that no such topological edge state exists for this lattice in the linear limit. Since this nonlinear state is also linearly unstable, the transient simulations reveal that the edge state sheds its energy. However, different from the nonlinearity-modified edge dynamics and similar to the dynamics of the unstable discrete Dirac soliton in Fig.~\ref{fig5}b, the edge state delocalization is accompanied by a localized wave that travels in the bulk. A similar phenomenon was also reported recently in a nonlinear SSH model of photonics~\cite{MS2021}. We conjecture that these belong to smaller energy and higher frequency soliton solutions near the optical band that is worth
exploring further.

\section{Conclusions \& Future Challenges}
In the present work, we have examined an SSH-type linear (dimer) system in the presence of an intersite nonlinearity of the $\beta$-FPUT type. We have leveraged our ability to control the linear band via a small parameter to develop a formal expansion in the vicinity of the band edge of the system. This, in turn, has led us to a variant of the nonlinear Dirac equations. We have used a sequence of linear and, subsequently, nonlinear (using polar coordinates) transformations to rewrite the relevant equations of motion. We observed that the equations simplify considerably in the limit of seeking the stationary nonlinear (continuum) wave. Eventually, the relevant coupled ODE problem is not only amenable to phase plane analysis, but it can also provide the soliton solution in closed analytical form. This, in turn, permits the computation of the associated conserved quantities (also discussed herein) ``at'' the solitonic solution. Armed with the knowledge of the coherent analytical structure, we then studied semi-infinite and finite-domain problems. There, we were able to show that a suitable adaptation of the soliton can be made to comply with concrete boundary conditions. This was sufficient (based on the bulk-boundary correspondence) to express the finite/semi-infinite domain edge states. We witness not only nonlinearity-modified topological edge states but also nonlinearity-induced edge states with no linear counterpart. The latter bifurcates from the bulk soliton solutions. We also examined the stability of the solitons and found that the deeper one goes into the gap, the more unstable the solutions. However, these instabilities were of an oscillatory type and tended to be weaker for large lattices, suggesting the stabilization in the infinite lattice limit. When the instability dynamics was explored, typically, we saw that a soliton-like wave was led to move within the lattice.

Naturally, this is only a first step towards the more systematic study of the lattices considered herein. One can envision numerous additional topics for future research. For instance, in the present work, we have limited our considerations to single stationary solitons. Yet, when instabilities arose, they often seemed to give rise to some propagating patterns spontaneously. It would be interesting to explore further whether such genuinely traveling structures exist (even if for isolated parameter values as, e.g., in the mass-dimer granular variant of~\cite{jayaprakash}) or not. 
It is interesting to point out in this context
that should such traveling wave solutions exist, the 
consideration of their momentum as a function of their 
speed would be worthwhile to consider in connection to their
stability, in line with classic studies along this vein, e.g.,
in~\cite{PhysRevLett.77.1193}.
%Multi-solitons and solitonic interactions (including as a function of their relative phase, a topic of particular, including experimental interest in the nonlinear Schr{\"o}dinger equation (NLS) realm~\cite{hulet}) constitute a natural extension of the present work. The collisions and the potential bound states of such structures within the lattice would be of interest. 
Furthermore, while we have constrained considerations to one-dimensional settings, generalizations to 2d lattices would be particularly interesting. This is due, among other things, to the fact that 2D nonlinear Dirac equations have been argued to have not only similarities but also intriguing differences from their NLS counterparts~\cite{PhysRevLett.116.214101}. This is both in terms of the stability of solitons and connection to the existence of vortical patterns. Such extensions are currently under consideration and will be reported in future publications.

\section*{ACKNOWLEDGMENTS}
R.C. acknowledges the funding support by the Science and Engineering
Research Board (SERB), India, through the Start-up Research Grant SRG/2022/001662.
P.G.K. acknowledges the support of the US National
Science Foundation under Grant Nos. DMS-2204702 and PHY-
2110030, as well as DMS-1809074.

\appendix
\section{Continuum vs. discrete dispersion}   
\label{AppendixA}
Here we verify that the dispersion relation in the continuum, as  
%given 
described by the PDEs in Eq.~\eqref{eq:pde_FPU}, 
%approximately 
captures fairly well the dispersion relation characterizing the discrete system given by Eq.~\eqref{eq:eom_FPU}. 
By substituting plane-wave solutions 
$u(z) = u_0 \exp[i(  \Tilde{\kappa}_c z - \Tilde{\omega}_c \tau)]$ and 
${v(z) = v_0 \exp[i(  \Tilde{\kappa}_c z - \Tilde{\omega}_c \tau)]}$ in  the linearized ($\mathrm{\Tilde{\Gamma}}=0$) Eq.~\eqref{eq:pde_FPU},  we get the dispersion relation for the continuum as

\begin{figure}[t!]
	\centering
	\includegraphics[width=\columnwidth]{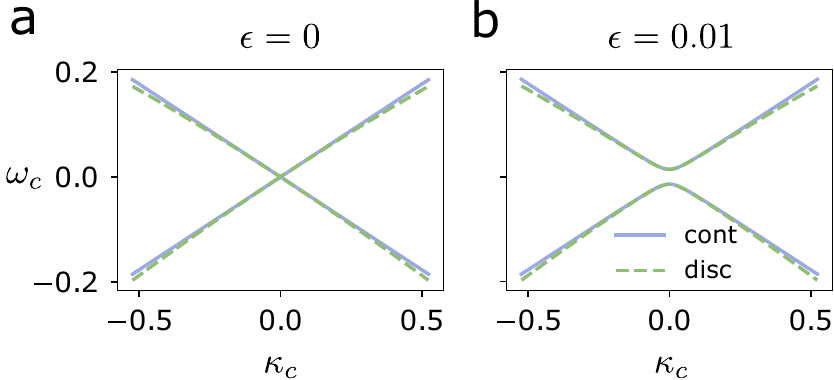}
	\caption{Comparison of continuum (solid) and discrete dispersion (dashed) for 
    (a) $\epsilon = 0$ and
		(b) $\epsilon = 0.01$ with $\Tilde{\gamma}=2$.}
	\label{figS1} 
\end{figure}

\begin{equation}
	\Tilde{\omega}_c = \pm \sqrt{{\Tilde{M}}^2 +  \Tilde{\kappa}_c^2 /8 }.
	\label{eq:PDE_dispersion0}
\end{equation}
Since this dispersion relation holds for the scaled coordinates, i.e., $z = \epsilon n$ and $\tau = \epsilon t $, an equivalent dispersion relation in the original coordinates $(n,t)$ would be:
\begin{equation}
	{\omega}_c = \pm \sqrt{{M}^2 +  {\kappa}_c^2 /8 },
	\label{eq:PDE_dispersion}
\end{equation}
where $ {\omega}_c = \epsilon \Tilde{\omega}_c $, 
$ {\kappa}_c = \epsilon \Tilde{\kappa}_c $, 
and $ M = \epsilon \Tilde{M}$
Similarly, we calculate the dispersion of the discrete system in Eq.~\eqref{eq:eom_FPU} as
\begin{equation}
	\omega_d = \sqrt{2 + \sqrt{(1-\gamma)^2 + (1+\gamma)^2  + 2(1-\gamma^2) \cos \kappa_d }}.
	\label{eq:discrete_dispersion}
\end{equation}
When using the ansatz in Eq.~\eqref{ansatz_FPU}, we know that plane-wave parameters are related as $\kappa_d = \abs{\pi + \kappa_c}$ and $\omega_d = \abs{\w + \omega_c}$. 
This means that the dispersion curve for the discrete chain in Eq.~\eqref{eq:discrete_dispersion}
has to be shifted in wavenumber and frequency to be compared to the dispersion in Eq.~\eqref{eq:PDE_dispersion}. We compare the two in Fig. \ref{figS1} and find a good match for the small band gap case of interest herein.

\section{Energy of the lattice vs. the power of NLD equations} 
\label{AppendixB}
\begin{figure}[t!]
	\centering
	\includegraphics[width=\columnwidth]{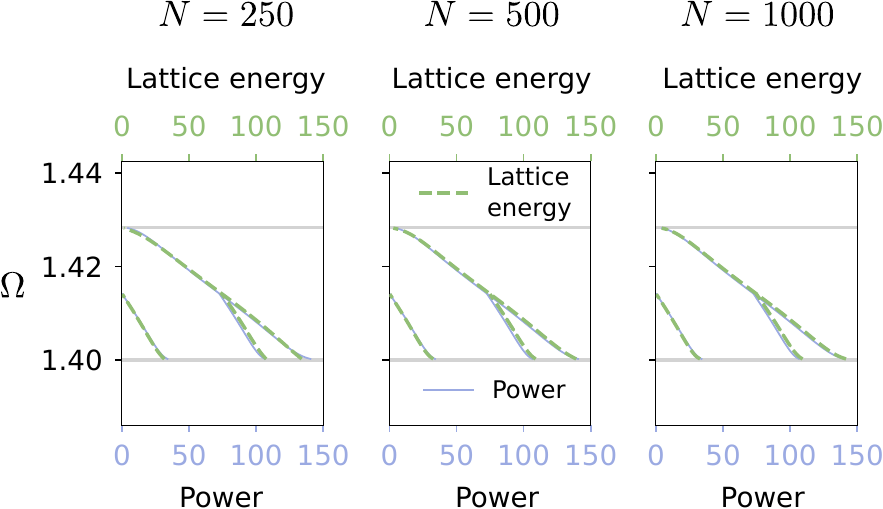}
	\caption{Comparison of NLD power and lattice energy for Dirac solitons and edge states in continuum. Three different lattice sizes are taken.}
	\label{figS2} 
\end{figure}

\begin{comment}
{\bf Two comments about this: the lattice energy,
I feel should be *defined inside the main text* and
should be used as equal to $\sum_n \varepsilon_n$ (by the
way there should be some subscript) so that $\varepsilon$
is defined as used in figures of the text. I am then
perplexed by the quasi-continuum approximation 
in that the quantity shown in the continuum limit is
not the continuum energy, nor is it clear that
it is conserved (as it stands). We need to discuss this.}
\end{comment}
For nonlinear periodic solutions (standing wave) at frequency $\w = \w_m + \epsilon \Tilde{\omega}$, let $|\xi|$ denote the amplitude of oscillations. We can then write the total energy of the lattice as the maximum potential energy
\begin{widetext}
\begin{eqnarray}
E_{\textrm{lattice}}(\Tilde{\omega})=\sum_{\text{unit cells}, \ n} \Big[  \frac{1}{4}  (1 &+&\gamma) (|\xi|_{2,n-1} - |\xi|_{1,n})^2 + \frac{1}{8}  \mathrm{\Gamma} (|\xi|_{2,n-1} - |\xi|_{1,n})^4 \nonumber \\
&+& \frac{1}{2} (1 -\gamma) (|\xi|_{1,n} - |\xi|_{2,n})^2  +  \frac{1}{4}  \mathrm{\Gamma}(|\xi|_{1,n} - |\xi|_{2,n})^4\nonumber \\
&+&  \frac{1}{4} (1+\gamma) (|\xi|_{2,n} - |\xi|_{1,n+1})^2 +  \frac{1}{8} \mathrm{\Gamma} (|\xi|_{2,n} - |\xi|_{1,n+1})^4\Big].
\end{eqnarray}
\end{widetext}
We then follow the same procedure described in Section~III and employ continuum approximation to reduce the equation in terms of the amplitude of ${\psi}_1$ and ${\psi_2}$, such that 
\begin{widetext}
\begin{eqnarray}
E_{\textrm{lattice}}(\Tilde{\omega})= \int_{-\infty}^{\infty} \left[     \frac{1}{4 {G} ^2 \sqrt{2}} \left(|{\psi}_1|^2 + |{\psi}_2|^2 \right)   +  \text{H.O.T.}   \right] ds,
\end{eqnarray}
\end{widetext}
where ${G}  = \sqrt{-3 \mathrm{{\Gamma}}/(4 \sqrt{2}})$. This equation is tantamount to the power of NLD [Eq.\eqref{eq:power}] for small $|{\psi}_1|$ and $|{\psi}_2|$. In Fig.~\ref{figS2}, we show the comparison of lattice energy and power for Dirac solitons and edge states that were found analytically for a continuum,
illustrating the very good agreement between the two.
%The excellent match between them indicates that lattice energy and NLD power (including a scale factor) tend to be synonymous for continuum solutions. 

\section{Boundary with $\theta(0) = 0$ ($U(0) = V(0)$)} 
\label{AppendixC}
For $\theta(0)= 0$, Eq.~\eqref{eq:theta_galilean} yields $s_0 =  0$.
Interestingly, this results in exactly the same profile as that of Dirac soliton; however, we take the right half for the semi-infinite domain $s \in [0, \infty)$. Since the Dirac soliton exists for the entire band gap, this edge state too exists for the entire band gap, i.e., $\Tilde{\omega} \in [-\Tilde{M}, \Tilde{M}]$. For the finite chain, this case corresponds to a \textit{free} end instead of a 
fixed one.

%Similarly, we look for solutions satisfying ${u(0)=-v(0)}$. This implies $\psi_1(0) = 0$, and then, $\theta(0)=\pi$. We thus deduce
%\begin{eqnarray}
%	s_0 =  - \frac{i}{\sqrt{M^2 - \omega^2}} \frac{\pi}{2},
%	\label{eq:free2}
%\end{eqnarray}
%which is purely imaginary for the entire band gap, i.e., $\omega \in [-M, M]$. Thus, it does not lead to any edge solution.

\section{Types of instabilities} 
\label{AppendixD}
\begin{figure*}[h!]
	\centering
	\includegraphics[width=\textwidth]{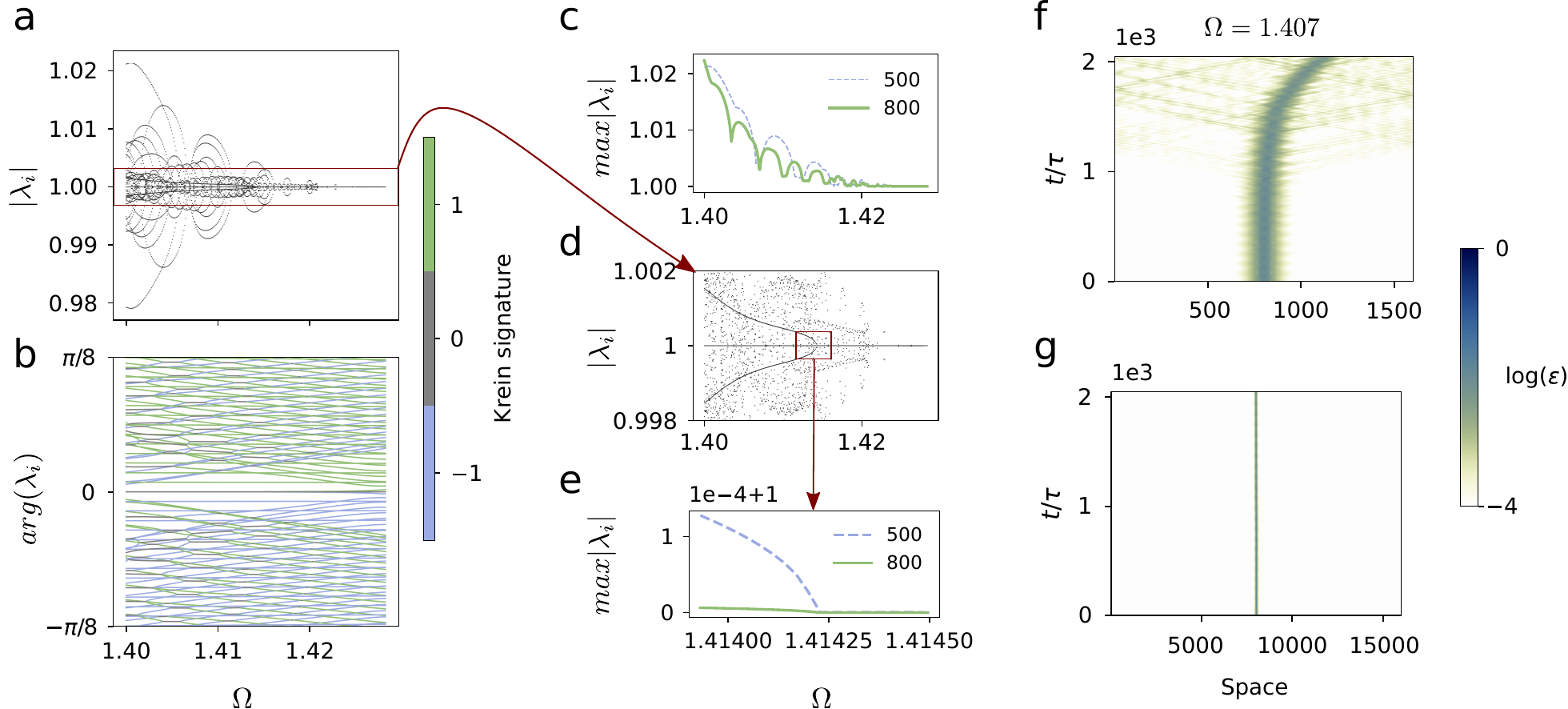}
	\caption{Stability of discrete Dirac soliton. (a) FM amplitude as a function of frequency inside the band gap for Dirac solitons. (b) The same for the FM phase. Colors denote the Krein signature of Floquet eigenmode. (c) FM amplitude decreases as the lattice length is increased, indicating the existence of finite-size instabilities. (d) Zoomed-in view of (a), indicating the emergence of a peculiar instability at the mid-gap frequency $\w_m=\sqrt{2}$. (e) This instability drastically diminishes with the increase in lattice length. (f) Spatial-temporal dynamics of Dirac soliton at $\w=1.407$ in a large lattice of $N=1500$. (g) The same as (f) but for an even larger lattice of $N=15000$. Colormap shows the energy density.}
	\label{figS3} 
\end{figure*}
Here we examine the instabilities of discrete Dirac solitons in more detail. In Figs.~\ref{figS3}a and~\ref{figS3}b, we show the amplitude and phase of FM for solitons inside the band gap. We observe that solitons become unstable, i.e., $|\lambda_i|>1$ for most of the frequencies below $\w=1.42$. Krein signature analysis~\cite{CXYKT2021} reveals that such instabilities are due to the collision of bulk spectrum in Fig.~\ref{figS3}b. Therefore, these are ``bulk-bulk'' or ``finite-size'' instabilities~\cite{CXYKT2021}. In Fig.~\ref{figS3}c, we show that such instabilities reduce with the increase of lattice size. This makes sense due to the existence of finite-size instabilities. In Fig.~\ref{figS3}d, we highlight 
%a peculiar 
the instability that emerges at the mid-gap frequency $\w=\sqrt{2}$. This instability is a result of coupling between the discrete Dirac soliton and the nonlinearity-induced edge state. As the length of the lattice is increased, this coupling is reduced, and thus, the instability drastically diminishes as shown in Fig.~\ref{figS3}e. We further confirm this by performing transient simulations on large lattices. We give the analytically obtained Dirac soliton solution at $\w=1.407$ as an initial condition to large lattices of size $N=1500$ and $N=15000$ in Figs.~\ref{figS3}f and~\ref{figS3}g. We observe  stable 
propagation of the breathing soliton in the larger lattice over
the course of the monitored time horizon, indicating the reduction of finite-size instabilities in such lattices.

\def\bibsection{\section*{References}} 

\def\bibsection{\section*{}} 

\bibliography{bibliography.bib}

\begin{thebibliography}{50}%
\makeatletter
\providecommand \@ifxundefined [1]{%
 \@ifx{#1\undefined}
}%
\providecommand \@ifnum [1]{%
 \ifnum #1\expandafter \@firstoftwo
 \else \expandafter \@secondoftwo
 \fi
}%
\providecommand \@ifx [1]{%
 \ifx #1\expandafter \@firstoftwo
 \else \expandafter \@secondoftwo
 \fi
}%
\providecommand \natexlab [1]{#1}%
\providecommand \enquote  [1]{``#1''}%
\providecommand \bibnamefont  [1]{#1}%
\providecommand \bibfnamefont [1]{#1}%
\providecommand \citenamefont [1]{#1}%
\providecommand \href@noop [0]{\@secondoftwo}%
\providecommand \href [0]{\begingroup \@sanitize@url \@href}%
\providecommand \@href[1]{\@@startlink{#1}\@@href}%
\providecommand \@@href[1]{\endgroup#1\@@endlink}%
\providecommand \@sanitize@url [0]{\catcode `\\12\catcode `\$12\catcode
  `\&12\catcode `\#12\catcode `\^12\catcode `\_12\catcode `\%12\relax}%
\providecommand \@@startlink[1]{}%
\providecommand \@@endlink[0]{}%
\providecommand \url  [0]{\begingroup\@sanitize@url \@url }%
\providecommand \@url [1]{\endgroup\@href {#1}{\urlprefix }}%
\providecommand \urlprefix  [0]{URL }%
\providecommand \Eprint [0]{\href }%
\providecommand \doibase [0]{https://doi.org/}%
\providecommand \selectlanguage [0]{\@gobble}%
\providecommand \bibinfo  [0]{\@secondoftwo}%
\providecommand \bibfield  [0]{\@secondoftwo}%
\providecommand \translation [1]{[#1]}%
\providecommand \BibitemOpen [0]{}%
\providecommand \bibitemStop [0]{}%
\providecommand \bibitemNoStop [0]{.\EOS\space}%
\providecommand \EOS [0]{\spacefactor3000\relax}%
\providecommand \BibitemShut  [1]{\csname bibitem#1\endcsname}%
\let\auto@bib@innerbib\@empty
%</preamble>
\bibitem [{\citenamefont {Fermi}\ \emph {et~al.}(1955)\citenamefont {Fermi},
  \citenamefont {Pasta},\ and\ \citenamefont {Ulam}}]{FPU55}%
  \BibitemOpen
  \bibfield  {author} {\bibinfo {author} {\bibfnamefont {E.}~\bibnamefont
  {Fermi}}, \bibinfo {author} {\bibfnamefont {J.}~\bibnamefont {Pasta}},\ and\
  \bibinfo {author} {\bibfnamefont {S.}~\bibnamefont {Ulam}},\ }\bibfield
  {title} {\bibinfo {title} {{Studies of Nonlinear Problems. I.}},\ }\href
  {https://www.osti.gov/biblio/4376203} {\bibfield  {journal} {\bibinfo
  {journal} {(Los Alamos National Laboratory, Los Alamos, NM, USA)}\ }\textbf
  {\bibinfo {volume} {Tech. Rep.}},\ \bibinfo {pages} {LA} (\bibinfo {year}
  {1955})}\BibitemShut {NoStop}%
\bibitem [{\citenamefont {Gallavotti}(2008)}]{FPUreview}%
  \BibitemOpen
  \bibfield  {author} {\bibinfo {author} {\bibfnamefont {G.}~\bibnamefont
  {Gallavotti}},\ }\href@noop {} {\emph {\bibinfo {title} {The
  Fermi--Pasta--Ulam Problem: A Status Report}}}\ (\bibinfo  {publisher}
  {Springer-Verlag, Berlin, Germany},\ \bibinfo {year} {2008})\BibitemShut
  {NoStop}%
\bibitem [{\citenamefont {Porter}\ \emph {et~al.}(2009)\citenamefont {Porter},
  \citenamefont {Zabusky}, \citenamefont {Hu},\ and\ \citenamefont
  {Campbell}}]{Zab5}%
  \BibitemOpen
  \bibfield  {author} {\bibinfo {author} {\bibfnamefont {M.}~\bibnamefont
  {Porter}}, \bibinfo {author} {\bibfnamefont {N.}~\bibnamefont {Zabusky}},
  \bibinfo {author} {\bibfnamefont {B.}~\bibnamefont {Hu}},\ and\ \bibinfo
  {author} {\bibfnamefont {D.}~\bibnamefont {Campbell}},\ }\bibfield  {title}
  {\bibinfo {title} {{Fermi, Pasta, Ulam and the Birth of Experimental
  Mathematics}},\ }\href
  {https://www.americanscientist.org/article/fermi-pasta-ulam-and-the-birth-of-experimental-mathematics}
  {\bibfield  {journal} {\bibinfo  {journal} {American Scientist}\ }\textbf
  {\bibinfo {volume} {97}},\ \bibinfo {pages} {214} (\bibinfo {year}
  {2009})}\BibitemShut {NoStop}%
\bibitem [{\citenamefont {Nesterenko}(2001)}]{Nester2001}%
  \BibitemOpen
  \bibfield  {author} {\bibinfo {author} {\bibfnamefont {V.~F.}\ \bibnamefont
  {Nesterenko}},\ }\href@noop {} {\emph {\bibinfo {title} {Dynamics of
  Heterogeneous Materials}}}\ (\bibinfo  {publisher} {Springer-Verlag},\
  \bibinfo {address} {Heidelberg, Germany},\ \bibinfo {year}
  {2001})\BibitemShut {NoStop}%
\bibitem [{\citenamefont {Chong}\ \emph {et~al.}(2017)\citenamefont {Chong},
  \citenamefont {Porter}, \citenamefont {Kevrekidis},\ and\ \citenamefont
  {Daraio}}]{gc_review}%
  \BibitemOpen
  \bibfield  {author} {\bibinfo {author} {\bibfnamefont {C.}~\bibnamefont
  {Chong}}, \bibinfo {author} {\bibfnamefont {M.~A.}\ \bibnamefont {Porter}},
  \bibinfo {author} {\bibfnamefont {P.~G.}\ \bibnamefont {Kevrekidis}},\ and\
  \bibinfo {author} {\bibfnamefont {C.}~\bibnamefont {Daraio}},\ }\bibfield
  {title} {\bibinfo {title} {{Nonlinear Coherent Structures in Granular
  Crystals}},\ }\href {https://dx.doi.org/10.1088/1361-648X/aa7672} {\bibfield
  {journal} {\bibinfo  {journal} {J. Phys.: Condens. Matter}\ }\textbf
  {\bibinfo {volume} {29}},\ \bibinfo {pages} {413003} (\bibinfo {year}
  {2017})}\BibitemShut {NoStop}%
\bibitem [{\citenamefont {Starosvetsky}\ \emph {et~al.}(2017)\citenamefont
  {Starosvetsky}, \citenamefont {Jayaprakash}, \citenamefont {Hasan},\ and\
  \citenamefont {Vakakis}}]{yuli_book}%
  \BibitemOpen
  \bibfield  {author} {\bibinfo {author} {\bibfnamefont {Y.}~\bibnamefont
  {Starosvetsky}}, \bibinfo {author} {\bibfnamefont {K.~R.}\ \bibnamefont
  {Jayaprakash}}, \bibinfo {author} {\bibfnamefont {M.~A.}\ \bibnamefont
  {Hasan}},\ and\ \bibinfo {author} {\bibfnamefont {A.~F.}\ \bibnamefont
  {Vakakis}},\ }\href@noop {} {\emph {\bibinfo {title} {Dynamics and Acoustics
  of Ordered Granular Media}}}\ (\bibinfo  {publisher} {World Scientific,
  Singapore},\ \bibinfo {year} {2017})\BibitemShut {NoStop}%
\bibitem [{\citenamefont {Theocharis}\ \emph {et~al.}(2013)\citenamefont
  {Theocharis}, \citenamefont {Boechler},\ and\ \citenamefont
  {Daraio}}]{granularBook}%
  \BibitemOpen
  \bibfield  {author} {\bibinfo {author} {\bibfnamefont {G.}~\bibnamefont
  {Theocharis}}, \bibinfo {author} {\bibfnamefont {N.}~\bibnamefont
  {Boechler}},\ and\ \bibinfo {author} {\bibfnamefont {C.}~\bibnamefont
  {Daraio}},\ }\href@noop {} {\emph {\bibinfo {title} {Nonlinear Periodic
  Phononic Structures and Granular Crystals}}}\ (\bibinfo  {publisher}
  {Springer Berlin Heidelberg},\ \bibinfo {address} {Berlin, Heidelberg},\
  \bibinfo {year} {2013})\ pp.\ \bibinfo {pages} {217--251}\BibitemShut
  {NoStop}%
\bibitem [{\citenamefont {Mehrem}\ \emph {et~al.}(2017)\citenamefont {Mehrem},
  \citenamefont {Jim\'enez}, \citenamefont {Salmer\'on-Contreras},
  \citenamefont {Garc\'{\i}a-Andr\'es}, \citenamefont {Garc\'{\i}a-Raffi},
  \citenamefont {Pic\'o},\ and\ \citenamefont
  {S\'anchez-Morcillo}}]{Mehrem2017}%
  \BibitemOpen
  \bibfield  {author} {\bibinfo {author} {\bibfnamefont {A.}~\bibnamefont
  {Mehrem}}, \bibinfo {author} {\bibfnamefont {N.}~\bibnamefont {Jim\'enez}},
  \bibinfo {author} {\bibfnamefont {L.~J.}\ \bibnamefont
  {Salmer\'on-Contreras}}, \bibinfo {author} {\bibfnamefont {X.}~\bibnamefont
  {Garc\'{\i}a-Andr\'es}}, \bibinfo {author} {\bibfnamefont {L.~M.}\
  \bibnamefont {Garc\'{\i}a-Raffi}}, \bibinfo {author} {\bibfnamefont
  {R.}~\bibnamefont {Pic\'o}},\ and\ \bibinfo {author} {\bibfnamefont {V.~J.}\
  \bibnamefont {S\'anchez-Morcillo}},\ }\bibfield  {title} {\bibinfo {title}
  {Nonlinear dispersive waves in repulsive lattices},\ }\href
  {https://link.aps.org/doi/10.1103/PhysRevE.96.012208} {\bibfield  {journal}
  {\bibinfo  {journal} {Phys. Rev. E}\ }\textbf {\bibinfo {volume} {96}},\
  \bibinfo {pages} {012208} (\bibinfo {year} {2017})}\BibitemShut {NoStop}%
\bibitem [{\citenamefont {Chong}\ \emph {et~al.}(2021)\citenamefont {Chong},
  \citenamefont {Wang}, \citenamefont {Mar{\'{e}}chal}, \citenamefont
  {Charalampidis}, \citenamefont {Moler{\'{o}}n}, \citenamefont
  {Mart{\'{\i}}nez}, \citenamefont {Porter}, \citenamefont {Kevrekidis},\ and\
  \citenamefont {Daraio}}]{Chong_2021}%
  \BibitemOpen
  \bibfield  {author} {\bibinfo {author} {\bibfnamefont {C.}~\bibnamefont
  {Chong}}, \bibinfo {author} {\bibfnamefont {Y.}~\bibnamefont {Wang}},
  \bibinfo {author} {\bibfnamefont {D.}~\bibnamefont {Mar{\'{e}}chal}},
  \bibinfo {author} {\bibfnamefont {E.~G.}\ \bibnamefont {Charalampidis}},
  \bibinfo {author} {\bibfnamefont {M.}~\bibnamefont {Moler{\'{o}}n}}, \bibinfo
  {author} {\bibfnamefont {A.~J.}\ \bibnamefont {Mart{\'{\i}}nez}}, \bibinfo
  {author} {\bibfnamefont {M.~A.}\ \bibnamefont {Porter}}, \bibinfo {author}
  {\bibfnamefont {P.~G.}\ \bibnamefont {Kevrekidis}},\ and\ \bibinfo {author}
  {\bibfnamefont {C.}~\bibnamefont {Daraio}},\ }\bibfield  {title} {\bibinfo
  {title} {Nonlinear localized modes in two-dimensional hexagonally-packed
  magnetic lattices},\ }\href {https://doi.org/10.1088/1367-2630/abdb6f}
  {\bibfield  {journal} {\bibinfo  {journal} {New J. Phys.}\ }\textbf {\bibinfo
  {volume} {23}},\ \bibinfo {pages} {043008} (\bibinfo {year}
  {2021})}\BibitemShut {NoStop}%
\bibitem [{\citenamefont {Yasuda}\ \emph {et~al.}(2019)\citenamefont {Yasuda},
  \citenamefont {Miyazawa}, \citenamefont {Charalampidis}, \citenamefont
  {Chong}, \citenamefont {Kevrekidis},\ and\ \citenamefont {Yang}}]{sci_adv}%
  \BibitemOpen
  \bibfield  {author} {\bibinfo {author} {\bibfnamefont {H.}~\bibnamefont
  {Yasuda}}, \bibinfo {author} {\bibfnamefont {Y.}~\bibnamefont {Miyazawa}},
  \bibinfo {author} {\bibfnamefont {E.~G.}\ \bibnamefont {Charalampidis}},
  \bibinfo {author} {\bibfnamefont {C.}~\bibnamefont {Chong}}, \bibinfo
  {author} {\bibfnamefont {P.~G.}\ \bibnamefont {Kevrekidis}},\ and\ \bibinfo
  {author} {\bibfnamefont {J.}~\bibnamefont {Yang}},\ }\bibfield  {title}
  {\bibinfo {title} {Origami-based impact mitigation via rarefaction solitary
  wave creation},\ }\href
  {https://www.science.org/doi/abs/10.1126/sciadv.aau2835} {\bibfield
  {journal} {\bibinfo  {journal} {Sci. Adv.}\ }\textbf {\bibinfo {volume}
  {5}},\ \bibinfo {pages} {eaau2835} (\bibinfo {year} {2019})}\BibitemShut
  {NoStop}%
\bibitem [{\citenamefont {Haddad}\ and\ \citenamefont {Carr}(2009)}]{Carr2009}%
  \BibitemOpen
  \bibfield  {author} {\bibinfo {author} {\bibfnamefont {L.}~\bibnamefont
  {Haddad}}\ and\ \bibinfo {author} {\bibfnamefont {L.}~\bibnamefont {Carr}},\
  }\bibfield  {title} {\bibinfo {title} {The nonlinear dirac equation in
  bose--einstein condensates: Foundation and symmetries},\ }\href
  {https://doi.org/https://doi.org/10.1016/j.physd.2009.02.001} {\bibfield
  {journal} {\bibinfo  {journal} {Physica D}\ }\textbf {\bibinfo {volume}
  {238}},\ \bibinfo {pages} {1413} (\bibinfo {year} {2009})},\ \bibinfo {note}
  {nonlinear Phenomena in Degenerate Quantum Gases}\BibitemShut {NoStop}%
\bibitem [{\citenamefont {Haddad}\ and\ \citenamefont {Carr}(2015)}]{l25}%
  \BibitemOpen
  \bibfield  {author} {\bibinfo {author} {\bibfnamefont {L.~H.}\ \bibnamefont
  {Haddad}}\ and\ \bibinfo {author} {\bibfnamefont {L.~D.}\ \bibnamefont
  {Carr}},\ }\bibfield  {title} {\bibinfo {title} {The nonlinear dirac equation
  in bose{\textendash}einstein condensates: vortex solutions and spectra in a
  weak harmonic trap},\ }\href {https://doi.org/10.1088/1367-2630/17/11/113011}
  {\bibfield  {journal} {\bibinfo  {journal} {New J. Phys.}\ }\textbf {\bibinfo
  {volume} {17}},\ \bibinfo {pages} {113011} (\bibinfo {year}
  {2015})}\BibitemShut {NoStop}%
\bibitem [{\citenamefont {Ablowitz}\ \emph {et~al.}(2009)\citenamefont
  {Ablowitz}, \citenamefont {Nixon},\ and\ \citenamefont {Zhu}}]{l30}%
  \BibitemOpen
  \bibfield  {author} {\bibinfo {author} {\bibfnamefont {M.~J.}\ \bibnamefont
  {Ablowitz}}, \bibinfo {author} {\bibfnamefont {S.~D.}\ \bibnamefont
  {Nixon}},\ and\ \bibinfo {author} {\bibfnamefont {Y.}~\bibnamefont {Zhu}},\
  }\bibfield  {title} {\bibinfo {title} {Conical diffraction in honeycomb
  lattices},\ }\href {https://doi.org/10.1103/PhysRevA.79.053830} {\bibfield
  {journal} {\bibinfo  {journal} {Phys. Rev. A}\ }\textbf {\bibinfo {volume}
  {79}},\ \bibinfo {pages} {053830} (\bibinfo {year} {2009})}\BibitemShut
  {NoStop}%
\bibitem [{\citenamefont {Ablowitz}\ and\ \citenamefont {Zhu}(2010)}]{l29}%
  \BibitemOpen
  \bibfield  {author} {\bibinfo {author} {\bibfnamefont {M.~J.}\ \bibnamefont
  {Ablowitz}}\ and\ \bibinfo {author} {\bibfnamefont {Y.}~\bibnamefont {Zhu}},\
  }\bibfield  {title} {\bibinfo {title} {Evolution of bloch-mode envelopes in
  two-dimensional generalized honeycomb lattices},\ }\href
  {https://doi.org/10.1103/PhysRevA.82.013840} {\bibfield  {journal} {\bibinfo
  {journal} {Phys. Rev. A}\ }\textbf {\bibinfo {volume} {82}},\ \bibinfo
  {pages} {013840} (\bibinfo {year} {2010})}\BibitemShut {NoStop}%
\bibitem [{\citenamefont {Cuevas-Maraver}\ \emph {et~al.}(2018)\citenamefont
  {Cuevas-Maraver}, \citenamefont {Boussa{\"i}d}, \citenamefont {Comech},
  \citenamefont {Lan}, \citenamefont {Kevrekidis},\ and\ \citenamefont
  {Saxena}}]{our_nab}%
  \BibitemOpen
  \bibfield  {author} {\bibinfo {author} {\bibfnamefont {J.}~\bibnamefont
  {Cuevas-Maraver}}, \bibinfo {author} {\bibfnamefont {N.}~\bibnamefont
  {Boussa{\"i}d}}, \bibinfo {author} {\bibfnamefont {A.}~\bibnamefont
  {Comech}}, \bibinfo {author} {\bibfnamefont {R.}~\bibnamefont {Lan}},
  \bibinfo {author} {\bibfnamefont {P.~G.}\ \bibnamefont {Kevrekidis}},\ and\
  \bibinfo {author} {\bibfnamefont {A.}~\bibnamefont {Saxena}},\ }\bibinfo
  {title} {Solitary waves in the nonlinear dirac equation},\ in\ \href
  {https://doi.org/10.1007/978-3-319-66766-9_4} {\emph {\bibinfo {booktitle}
  {Nonlinear Systems, Vol. 1: Mathematical Theory and Computational
  Methods}}},\ \bibinfo {editor} {edited by\ \bibinfo {editor} {\bibfnamefont
  {V.}~\bibnamefont {Carmona}}, \bibinfo {editor} {\bibfnamefont
  {J.}~\bibnamefont {Cuevas-Maraver}}, \bibinfo {editor} {\bibfnamefont
  {F.}~\bibnamefont {Fern{\'a}ndez-S{\'a}nchez}},\ and\ \bibinfo {editor}
  {\bibfnamefont {E.}~\bibnamefont {Garc{\'i}a-Medina}}}\ (\bibinfo
  {publisher} {Springer International Publishing},\ \bibinfo {address} {Cham},\
  \bibinfo {year} {2018})\ pp.\ \bibinfo {pages} {89--143}\BibitemShut
  {NoStop}%
\bibitem [{\citenamefont {Boussa{\"i}d}\ and\ \citenamefont
  {Comech}(2019)}]{nabil2}%
  \BibitemOpen
  \bibfield  {author} {\bibinfo {author} {\bibfnamefont {N.}~\bibnamefont
  {Boussa{\"i}d}}\ and\ \bibinfo {author} {\bibfnamefont {A.}~\bibnamefont
  {Comech}},\ }\href@noop {} {\emph {\bibinfo {title} {Nonlinear Dirac
  Equation: Spectral Stability of Solitary Waves}}}\ (\bibinfo  {publisher}
  {American Mathematical Society},\ \bibinfo {address} {Providence, RI},\
  \bibinfo {year} {2019})\BibitemShut {NoStop}%
\bibitem [{\citenamefont {Hasan}\ and\ \citenamefont {Kane}(2010)}]{Hasan2010}%
  \BibitemOpen
  \bibfield  {author} {\bibinfo {author} {\bibfnamefont {M.~Z.}\ \bibnamefont
  {Hasan}}\ and\ \bibinfo {author} {\bibfnamefont {C.~L.}\ \bibnamefont
  {Kane}},\ }\bibfield  {title} {\bibinfo {title} {{Colloquium : Topological
  insulators}},\ }\href {https://doi.org/10.1103/RevModPhys.82.3045} {\bibfield
   {journal} {\bibinfo  {journal} {Rev. Mod. Phys.}\ }\textbf {\bibinfo
  {volume} {82}},\ \bibinfo {pages} {3045} (\bibinfo {year}
  {2010})}\BibitemShut {NoStop}%
\bibitem [{\citenamefont {Cooper}\ \emph {et~al.}(2019)\citenamefont {Cooper},
  \citenamefont {Dalibard},\ and\ \citenamefont {Spielman}}]{Cooper2019}%
  \BibitemOpen
  \bibfield  {author} {\bibinfo {author} {\bibfnamefont {N.~R.}\ \bibnamefont
  {Cooper}}, \bibinfo {author} {\bibfnamefont {J.}~\bibnamefont {Dalibard}},\
  and\ \bibinfo {author} {\bibfnamefont {I.~B.}\ \bibnamefont {Spielman}},\
  }\bibfield  {title} {\bibinfo {title} {Topological bands for ultracold
  atoms},\ }\href {https://doi.org/10.1103/RevModPhys.91.015005} {\bibfield
  {journal} {\bibinfo  {journal} {Rev. Mod. Phys.}\ }\textbf {\bibinfo {volume}
  {91}},\ \bibinfo {pages} {015005} (\bibinfo {year} {2019})}\BibitemShut
  {NoStop}%
\bibitem [{\citenamefont {Ozawa}\ \emph {et~al.}(2019)\citenamefont {Ozawa},
  \citenamefont {Price}, \citenamefont {Amo}, \citenamefont {Goldman},
  \citenamefont {Hafezi}, \citenamefont {Lu}, \citenamefont {Rechtsman},
  \citenamefont {Schuster}, \citenamefont {Simon}, \citenamefont {Zilberberg},\
  and\ \citenamefont {Carusotto}}]{Ozawa2019}%
  \BibitemOpen
  \bibfield  {author} {\bibinfo {author} {\bibfnamefont {T.}~\bibnamefont
  {Ozawa}}, \bibinfo {author} {\bibfnamefont {H.~M.}\ \bibnamefont {Price}},
  \bibinfo {author} {\bibfnamefont {A.}~\bibnamefont {Amo}}, \bibinfo {author}
  {\bibfnamefont {N.}~\bibnamefont {Goldman}}, \bibinfo {author} {\bibfnamefont
  {M.}~\bibnamefont {Hafezi}}, \bibinfo {author} {\bibfnamefont
  {L.}~\bibnamefont {Lu}}, \bibinfo {author} {\bibfnamefont {M.~C.}\
  \bibnamefont {Rechtsman}}, \bibinfo {author} {\bibfnamefont {D.}~\bibnamefont
  {Schuster}}, \bibinfo {author} {\bibfnamefont {J.}~\bibnamefont {Simon}},
  \bibinfo {author} {\bibfnamefont {O.}~\bibnamefont {Zilberberg}},\ and\
  \bibinfo {author} {\bibfnamefont {I.}~\bibnamefont {Carusotto}},\ }\bibfield
  {title} {\bibinfo {title} {{Topological photonics}},\ }\href
  {https://doi.org/10.1103/RevModPhys.91.015006} {\bibfield  {journal}
  {\bibinfo  {journal} {Rev. Mod. Phys.}\ }\textbf {\bibinfo {volume} {91}},\
  \bibinfo {pages} {015006} (\bibinfo {year} {2019})}\BibitemShut {NoStop}%
\bibitem [{\citenamefont {S{\"{u}}sstrunk}\ and\ \citenamefont
  {Huber}(2016)}]{Susstrunk2016}%
  \BibitemOpen
  \bibfield  {author} {\bibinfo {author} {\bibfnamefont {R.}~\bibnamefont
  {S{\"{u}}sstrunk}}\ and\ \bibinfo {author} {\bibfnamefont {S.~D.}\
  \bibnamefont {Huber}},\ }\bibfield  {title} {\bibinfo {title}
  {{Classification of topological phonons in linear mechanical
  metamaterials}},\ }\href {https://doi.org/10.1073/pnas.1605462113} {\bibfield
   {journal} {\bibinfo  {journal} {Proc. Natl. Acad. Sci. USA}\ }\textbf
  {\bibinfo {volume} {113}},\ \bibinfo {pages} {E4767} (\bibinfo {year}
  {2016})}\BibitemShut {NoStop}%
\bibitem [{\citenamefont {Ma}\ \emph {et~al.}(2019)\citenamefont {Ma},
  \citenamefont {Xiao},\ and\ \citenamefont {Chan}}]{Ma2019}%
  \BibitemOpen
  \bibfield  {author} {\bibinfo {author} {\bibfnamefont {G.}~\bibnamefont
  {Ma}}, \bibinfo {author} {\bibfnamefont {M.}~\bibnamefont {Xiao}},\ and\
  \bibinfo {author} {\bibfnamefont {C.~T.}\ \bibnamefont {Chan}},\ }\bibfield
  {title} {\bibinfo {title} {{Topological phases in acoustic and mechanical
  systems}},\ }\href {http://www.nature.com/articles/s42254-019-0030-x}
  {\bibfield  {journal} {\bibinfo  {journal} {Nat. Rev. Phys.}\ }\textbf
  {\bibinfo {volume} {1}},\ \bibinfo {pages} {281} (\bibinfo {year}
  {2019})}\BibitemShut {NoStop}%
\bibitem [{\citenamefont {Bernevig}\ and\ \citenamefont
  {Hughes}(2013)}]{Bernevig2013}%
  \BibitemOpen
  \bibfield  {author} {\bibinfo {author} {\bibfnamefont {B.~A.}\ \bibnamefont
  {Bernevig}}\ and\ \bibinfo {author} {\bibfnamefont {T.~L.}\ \bibnamefont
  {Hughes}},\ }\href {http://www.jstor.org/stable/j.ctt19cc2gc} {\emph
  {\bibinfo {title} {Topological Insulators and Topological Superconductors}}}\
  (\bibinfo  {publisher} {Princeton University Press},\ \bibinfo {year}
  {2013})\BibitemShut {NoStop}%
\bibitem [{\citenamefont {Kane}\ and\ \citenamefont {Mele}(2005)}]{Kane2005}%
  \BibitemOpen
  \bibfield  {author} {\bibinfo {author} {\bibfnamefont {C.~L.}\ \bibnamefont
  {Kane}}\ and\ \bibinfo {author} {\bibfnamefont {E.~J.}\ \bibnamefont
  {Mele}},\ }\bibfield  {title} {\bibinfo {title} {Quantum spin hall effect in
  graphene},\ }\href {https://doi.org/10.1103/PhysRevLett.95.226801} {\bibfield
   {journal} {\bibinfo  {journal} {Phys. Rev. Lett.}\ }\textbf {\bibinfo
  {volume} {95}},\ \bibinfo {pages} {226801} (\bibinfo {year}
  {2005})}\BibitemShut {NoStop}%
\bibitem [{\citenamefont {Wan}\ \emph {et~al.}(2011)\citenamefont {Wan},
  \citenamefont {Turner}, \citenamefont {Vishwanath},\ and\ \citenamefont
  {Savrasov}}]{Wan2011}%
  \BibitemOpen
  \bibfield  {author} {\bibinfo {author} {\bibfnamefont {X.}~\bibnamefont
  {Wan}}, \bibinfo {author} {\bibfnamefont {A.~M.}\ \bibnamefont {Turner}},
  \bibinfo {author} {\bibfnamefont {A.}~\bibnamefont {Vishwanath}},\ and\
  \bibinfo {author} {\bibfnamefont {S.~Y.}\ \bibnamefont {Savrasov}},\
  }\bibfield  {title} {\bibinfo {title} {Topological semimetal and fermi-arc
  surface states in the electronic structure of pyrochlore iridates},\ }\href
  {https://doi.org/10.1103/PhysRevB.83.205101} {\bibfield  {journal} {\bibinfo
  {journal} {Phys. Rev. B}\ }\textbf {\bibinfo {volume} {83}},\ \bibinfo
  {pages} {205101} (\bibinfo {year} {2011})}\BibitemShut {NoStop}%
\bibitem [{\citenamefont {Benalcazar}\ \emph {et~al.}(2017)\citenamefont
  {Benalcazar}, \citenamefont {Bernevig},\ and\ \citenamefont
  {Hughes}}]{Benalcazar2017}%
  \BibitemOpen
  \bibfield  {author} {\bibinfo {author} {\bibfnamefont {W.~A.}\ \bibnamefont
  {Benalcazar}}, \bibinfo {author} {\bibfnamefont {B.~A.}\ \bibnamefont
  {Bernevig}},\ and\ \bibinfo {author} {\bibfnamefont {T.~L.}\ \bibnamefont
  {Hughes}},\ }\bibfield  {title} {\bibinfo {title} {Quantized electric
  multipole insulators},\ }\href {https://doi.org/10.1126/science.aah6442}
  {\bibfield  {journal} {\bibinfo  {journal} {Science}\ }\textbf {\bibinfo
  {volume} {357}},\ \bibinfo {pages} {61} (\bibinfo {year} {2017})}\BibitemShut
  {NoStop}%
\bibitem [{\citenamefont {Flach}\ and\ \citenamefont
  {Gorbach}(2008)}]{Flach2008}%
  \BibitemOpen
  \bibfield  {author} {\bibinfo {author} {\bibfnamefont {S.}~\bibnamefont
  {Flach}}\ and\ \bibinfo {author} {\bibfnamefont {A.~V.}\ \bibnamefont
  {Gorbach}},\ }\bibfield  {title} {\bibinfo {title} {{Discrete breathers —
  Advances in theory and applications}},\ }\href
  {https://doi.org/10.1016/j.physrep.2008.05.002} {\bibfield  {journal}
  {\bibinfo  {journal} {Phys. Rep.}\ }\textbf {\bibinfo {volume} {467}},\
  \bibinfo {pages} {1} (\bibinfo {year} {2008})}\BibitemShut {NoStop}%
\bibitem [{\citenamefont {Aubry}(2006)}]{Aubry2006}%
  \BibitemOpen
  \bibfield  {author} {\bibinfo {author} {\bibfnamefont {S.}~\bibnamefont
  {Aubry}},\ }\bibfield  {title} {\bibinfo {title} {{Discrete Breathers:
  Localization and transfer of energy in discrete Hamiltonian nonlinear
  systems}},\ }\href {https://doi.org/10.1016/j.physd.2005.12.020} {\bibfield
  {journal} {\bibinfo  {journal} {Physica D}\ }\textbf {\bibinfo {volume}
  {216}},\ \bibinfo {pages} {1} (\bibinfo {year} {2006})}\BibitemShut {NoStop}%
\bibitem [{\citenamefont {Livi}\ \emph {et~al.}(1997)\citenamefont {Livi},
  \citenamefont {Spicci},\ and\ \citenamefont {MacKay}}]{Livi_1997}%
  \BibitemOpen
  \bibfield  {author} {\bibinfo {author} {\bibfnamefont {R.}~\bibnamefont
  {Livi}}, \bibinfo {author} {\bibfnamefont {M.}~\bibnamefont {Spicci}},\ and\
  \bibinfo {author} {\bibfnamefont {R.~S.}\ \bibnamefont {MacKay}},\ }\bibfield
   {title} {\bibinfo {title} {Breathers on a diatomic {FPU} chain},\ }\href
  {https://doi.org/10.1088/0951-7715/10/6/003} {\bibfield  {journal} {\bibinfo
  {journal} {Nonlinearity}\ }\textbf {\bibinfo {volume} {10}},\ \bibinfo
  {pages} {1421} (\bibinfo {year} {1997})}\BibitemShut {NoStop}%
\bibitem [{\citenamefont {Maniadis}\ \emph {et~al.}(2003)\citenamefont
  {Maniadis}, \citenamefont {Zolotaryuk},\ and\ \citenamefont
  {Tsironis}}]{maniadis}%
  \BibitemOpen
  \bibfield  {author} {\bibinfo {author} {\bibfnamefont {P.}~\bibnamefont
  {Maniadis}}, \bibinfo {author} {\bibfnamefont {A.~V.}\ \bibnamefont
  {Zolotaryuk}},\ and\ \bibinfo {author} {\bibfnamefont {G.~P.}\ \bibnamefont
  {Tsironis}},\ }\bibfield  {title} {\bibinfo {title} {Existence and stability
  of discrete gap breathers in a diatomic $\ensuremath{\beta}$ fermi-pasta-ulam
  chain},\ }\href {https://doi.org/10.1103/PhysRevE.67.046612} {\bibfield
  {journal} {\bibinfo  {journal} {Phys. Rev. E}\ }\textbf {\bibinfo {volume}
  {67}},\ \bibinfo {pages} {046612} (\bibinfo {year} {2003})}\BibitemShut
  {NoStop}%
\bibitem [{\citenamefont {James}\ and\ \citenamefont
  {Noble}(2004)}]{JAMES2004124}%
  \BibitemOpen
  \bibfield  {author} {\bibinfo {author} {\bibfnamefont {G.}~\bibnamefont
  {James}}\ and\ \bibinfo {author} {\bibfnamefont {P.}~\bibnamefont {Noble}},\
  }\bibfield  {title} {\bibinfo {title} {Breathers on diatomic
  fermi–pasta–ulam lattices},\ }\href
  {https://doi.org/https://doi.org/10.1016/j.physd.2004.05.005} {\bibfield
  {journal} {\bibinfo  {journal} {Physica D}\ }\textbf {\bibinfo {volume}
  {196}},\ \bibinfo {pages} {124} (\bibinfo {year} {2004})}\BibitemShut
  {NoStop}%
\bibitem [{\citenamefont {Su}\ \emph {et~al.}(1979)\citenamefont {Su},
  \citenamefont {Schrieffer},\ and\ \citenamefont {Heeger}}]{Su1979}%
  \BibitemOpen
  \bibfield  {author} {\bibinfo {author} {\bibfnamefont {W.~P.}\ \bibnamefont
  {Su}}, \bibinfo {author} {\bibfnamefont {J.~R.}\ \bibnamefont {Schrieffer}},\
  and\ \bibinfo {author} {\bibfnamefont {A.~J.}\ \bibnamefont {Heeger}},\
  }\bibfield  {title} {\bibinfo {title} {{Solitons in Polyacetylene}},\ }\href
  {https://doi.org/10.1103/PhysRevLett.42.1698} {\bibfield  {journal} {\bibinfo
   {journal} {Phys. Rev. Lett.}\ }\textbf {\bibinfo {volume} {42}},\ \bibinfo
  {pages} {1698} (\bibinfo {year} {1979})}\BibitemShut {NoStop}%
\bibitem [{\citenamefont {Ma}\ and\ \citenamefont {Susanto}(2021)}]{MS2021}%
  \BibitemOpen
  \bibfield  {author} {\bibinfo {author} {\bibfnamefont {Y.-P.}\ \bibnamefont
  {Ma}}\ and\ \bibinfo {author} {\bibfnamefont {H.}~\bibnamefont {Susanto}},\
  }\bibfield  {title} {\bibinfo {title} {Topological edge solitons and their
  stability in a nonlinear su-schrieffer-heeger model},\ }\href
  {https://doi.org/10.1103/PhysRevE.104.054206} {\bibfield  {journal} {\bibinfo
   {journal} {Phys. Rev. E}\ }\textbf {\bibinfo {volume} {104}},\ \bibinfo
  {pages} {054206} (\bibinfo {year} {2021})}\BibitemShut {NoStop}%
\bibitem [{\citenamefont {Chaunsali}\ \emph {et~al.}(2021)\citenamefont
  {Chaunsali}, \citenamefont {Xu}, \citenamefont {Yang}, \citenamefont
  {Kevrekidis},\ and\ \citenamefont {Theocharis}}]{CXYKT2021}%
  \BibitemOpen
  \bibfield  {author} {\bibinfo {author} {\bibfnamefont {R.}~\bibnamefont
  {Chaunsali}}, \bibinfo {author} {\bibfnamefont {H.}~\bibnamefont {Xu}},
  \bibinfo {author} {\bibfnamefont {J.}~\bibnamefont {Yang}}, \bibinfo {author}
  {\bibfnamefont {P.~G.}\ \bibnamefont {Kevrekidis}},\ and\ \bibinfo {author}
  {\bibfnamefont {G.}~\bibnamefont {Theocharis}},\ }\bibfield  {title}
  {\bibinfo {title} {Stability of topological edge states under strong
  nonlinear effects},\ }\href {https://doi.org/10.1103/PhysRevB.103.024106}
  {\bibfield  {journal} {\bibinfo  {journal} {Phys. Rev. B}\ }\textbf {\bibinfo
  {volume} {103}},\ \bibinfo {pages} {024106} (\bibinfo {year}
  {2021})}\BibitemShut {NoStop}%
\bibitem [{\citenamefont {Hofstrand}\ \emph {et~al.}(2023)\citenamefont
  {Hofstrand}, \citenamefont {Li},\ and\ \citenamefont
  {Weinstein}}]{hofstrand}%
  \BibitemOpen
  \bibfield  {author} {\bibinfo {author} {\bibfnamefont {A.}~\bibnamefont
  {Hofstrand}}, \bibinfo {author} {\bibfnamefont {H.}~\bibnamefont {Li}},\ and\
  \bibinfo {author} {\bibfnamefont {M.~I.}\ \bibnamefont {Weinstein}},\
  }\bibfield  {title} {\bibinfo {title} {Discrete breathers of nonlinear dimer
  lattices: Bridging the anti-continuous and continuous limits},\ }\href
  {https://doi.org/10.1007/s00332-023-09909-x} {\bibfield  {journal} {\bibinfo
  {journal} {J Nonlinear Sci}\ }\textbf {\bibinfo {volume} {33}},\ \bibinfo
  {pages} {59} (\bibinfo {year} {2023})}\BibitemShut {NoStop}%
\bibitem [{\citenamefont {Smirnova}\ \emph {et~al.}(2019)\citenamefont
  {Smirnova}, \citenamefont {Smirnov}, \citenamefont {Leykam},\ and\
  \citenamefont {Kivshar}}]{SSLK2019}%
  \BibitemOpen
  \bibfield  {author} {\bibinfo {author} {\bibfnamefont {D.~A.}\ \bibnamefont
  {Smirnova}}, \bibinfo {author} {\bibfnamefont {L.~A.}\ \bibnamefont
  {Smirnov}}, \bibinfo {author} {\bibfnamefont {D.}~\bibnamefont {Leykam}},\
  and\ \bibinfo {author} {\bibfnamefont {Y.~S.}\ \bibnamefont {Kivshar}},\
  }\bibfield  {title} {\bibinfo {title} {Topological edge states and gap
  solitons in the nonlinear dirac model},\ }\href
  {https://doi.org/https://doi.org/10.1002/lpor.201900223} {\bibfield
  {journal} {\bibinfo  {journal} {Laser \& Photonics Rev.}\ }\textbf {\bibinfo
  {volume} {13}},\ \bibinfo {pages} {1900223} (\bibinfo {year}
  {2019})}\BibitemShut {NoStop}%
\bibitem [{\citenamefont {Berkolaiko}\ \emph {et~al.}(2015)\citenamefont
  {Berkolaiko}, \citenamefont {Comech},\ and\ \citenamefont
  {Sukhtayev}}]{Berkolaiko_2015}%
  \BibitemOpen
  \bibfield  {author} {\bibinfo {author} {\bibfnamefont {G.}~\bibnamefont
  {Berkolaiko}}, \bibinfo {author} {\bibfnamefont {A.}~\bibnamefont {Comech}},\
  and\ \bibinfo {author} {\bibfnamefont {A.}~\bibnamefont {Sukhtayev}},\
  }\bibfield  {title} {\bibinfo {title} {Vakhitov{\textendash}kolokolov and
  energy vanishing conditions for linear instability of solitary waves in
  models of classical self-interacting spinor fields},\ }\href
  {https://doi.org/10.1088/0951-7715/28/3/577} {\bibfield  {journal} {\bibinfo
  {journal} {Nonlinearity}\ }\textbf {\bibinfo {volume} {28}},\ \bibinfo
  {pages} {577} (\bibinfo {year} {2015})}\BibitemShut {NoStop}%
\bibitem [{\citenamefont {Chaunsali}\ and\ \citenamefont
  {Theocharis}(2019)}]{CT2019}%
  \BibitemOpen
  \bibfield  {author} {\bibinfo {author} {\bibfnamefont {R.}~\bibnamefont
  {Chaunsali}}\ and\ \bibinfo {author} {\bibfnamefont {G.}~\bibnamefont
  {Theocharis}},\ }\bibfield  {title} {\bibinfo {title} {Self-induced
  topological transition in phononic crystals by nonlinearity management},\
  }\href {https://doi.org/10.1103/PhysRevB.100.014302} {\bibfield  {journal}
  {\bibinfo  {journal} {Phys. Rev. B}\ }\textbf {\bibinfo {volume} {100}},\
  \bibinfo {pages} {014302} (\bibinfo {year} {2019})}\BibitemShut {NoStop}%
\bibitem [{\citenamefont {Kivshar}(1992)}]{Kivshar1992}%
  \BibitemOpen
  \bibfield  {author} {\bibinfo {author} {\bibfnamefont {Y.~S.}\ \bibnamefont
  {Kivshar}},\ }\bibfield  {title} {\bibinfo {title} {Class of localized
  structures in nonlinear lattices},\ }\href
  {https://link.aps.org/doi/10.1103/PhysRevB.46.8652} {\bibfield  {journal}
  {\bibinfo  {journal} {Phys. Rev. B}\ }\textbf {\bibinfo {volume} {46}},\
  \bibinfo {pages} {8652} (\bibinfo {year} {1992})}\BibitemShut {NoStop}%
\bibitem [{\citenamefont {Chubykalo}\ \emph {et~al.}(1993)\citenamefont
  {Chubykalo}, \citenamefont {Kovalev},\ and\ \citenamefont
  {Usatenko}}]{Chubykalo1993b}%
  \BibitemOpen
  \bibfield  {author} {\bibinfo {author} {\bibfnamefont {O.~A.}\ \bibnamefont
  {Chubykalo}}, \bibinfo {author} {\bibfnamefont {A.~S.}\ \bibnamefont
  {Kovalev}},\ and\ \bibinfo {author} {\bibfnamefont {O.~V.}\ \bibnamefont
  {Usatenko}},\ }\bibfield  {title} {\bibinfo {title} {Dynamical solitons in a
  one-dimensional nonlinear diatomic chain},\ }\href
  {https://link.aps.org/doi/10.1103/PhysRevB.47.3153} {\bibfield  {journal}
  {\bibinfo  {journal} {Phys. Rev. B}\ }\textbf {\bibinfo {volume} {47}},\
  \bibinfo {pages} {3153} (\bibinfo {year} {1993})}\BibitemShut {NoStop}%
\bibitem [{\citenamefont {Alexeeva}\ \emph {et~al.}(2019)\citenamefont
  {Alexeeva}, \citenamefont {Barashenkov},\ and\ \citenamefont
  {Saxena}}]{ALEXEEVA2019198}%
  \BibitemOpen
  \bibfield  {author} {\bibinfo {author} {\bibfnamefont {N.}~\bibnamefont
  {Alexeeva}}, \bibinfo {author} {\bibfnamefont {I.}~\bibnamefont
  {Barashenkov}},\ and\ \bibinfo {author} {\bibfnamefont {A.}~\bibnamefont
  {Saxena}},\ }\bibfield  {title} {\bibinfo {title} {Spinor solitons and their
  pt-symmetric offspring},\ }\href
  {https://doi.org/https://doi.org/10.1016/j.aop.2018.11.010} {\bibfield
  {journal} {\bibinfo  {journal} {Annals of Physics}\ }\textbf {\bibinfo
  {volume} {403}},\ \bibinfo {pages} {198} (\bibinfo {year}
  {2019})}\BibitemShut {NoStop}%
\bibitem [{\citenamefont {Huang}\ and\ \citenamefont {Hu}(1998)}]{Huang1998}%
  \BibitemOpen
  \bibfield  {author} {\bibinfo {author} {\bibfnamefont {G.}~\bibnamefont
  {Huang}}\ and\ \bibinfo {author} {\bibfnamefont {B.}~\bibnamefont {Hu}},\
  }\bibfield  {title} {\bibinfo {title} {Asymmetric gap soliton modes in
  diatomic lattices with cubic and quartic nonlinearity},\ }\href
  {https://link.aps.org/doi/10.1103/PhysRevB.57.5746} {\bibfield  {journal}
  {\bibinfo  {journal} {Phys. Rev. B}\ }\textbf {\bibinfo {volume} {57}},\
  \bibinfo {pages} {5746} (\bibinfo {year} {1998})}\BibitemShut {NoStop}%
\bibitem [{\citenamefont {Hu}\ \emph {et~al.}(2000)\citenamefont {Hu},
  \citenamefont {Huang},\ and\ \citenamefont {Velarde}}]{Huang2000}%
  \BibitemOpen
  \bibfield  {author} {\bibinfo {author} {\bibfnamefont {B.}~\bibnamefont
  {Hu}}, \bibinfo {author} {\bibfnamefont {G.}~\bibnamefont {Huang}},\ and\
  \bibinfo {author} {\bibfnamefont {M.~G.}\ \bibnamefont {Velarde}},\
  }\bibfield  {title} {\bibinfo {title} {Dynamics of coupled gap solitons in
  diatomic lattices with cubic and quartic nonlinearities},\ }\href
  {https://link.aps.org/doi/10.1103/PhysRevE.62.2827} {\bibfield  {journal}
  {\bibinfo  {journal} {Phys. Rev. E}\ }\textbf {\bibinfo {volume} {62}},\
  \bibinfo {pages} {2827} (\bibinfo {year} {2000})}\BibitemShut {NoStop}%
\bibitem [{\citenamefont {Mills}\ and\ \citenamefont
  {Trullinger}(1987)}]{Mills1987}%
  \BibitemOpen
  \bibfield  {author} {\bibinfo {author} {\bibfnamefont {D.~L.}\ \bibnamefont
  {Mills}}\ and\ \bibinfo {author} {\bibfnamefont {S.~E.}\ \bibnamefont
  {Trullinger}},\ }\bibfield  {title} {\bibinfo {title} {Gap solitons in
  nonlinear periodic structures},\ }\href
  {https://link.aps.org/doi/10.1103/PhysRevB.36.947} {\bibfield  {journal}
  {\bibinfo  {journal} {Phys. Rev. B}\ }\textbf {\bibinfo {volume} {36}},\
  \bibinfo {pages} {947} (\bibinfo {year} {1987})}\BibitemShut {NoStop}%
\bibitem [{\citenamefont {Aceves}\ and\ \citenamefont
  {Wabnitz}(1989)}]{ACEVES198937}%
  \BibitemOpen
  \bibfield  {author} {\bibinfo {author} {\bibfnamefont {A.}~\bibnamefont
  {Aceves}}\ and\ \bibinfo {author} {\bibfnamefont {S.}~\bibnamefont
  {Wabnitz}},\ }\bibfield  {title} {\bibinfo {title} {Self-induced transparency
  solitons in nonlinear refractive periodic media},\ }\href
  {https://doi.org/https://doi.org/10.1016/0375-9601(89)90441-6} {\bibfield
  {journal} {\bibinfo  {journal} {Phys. Lett. A}\ }\textbf {\bibinfo {volume}
  {141}},\ \bibinfo {pages} {37} (\bibinfo {year} {1989})}\BibitemShut
  {NoStop}%
\bibitem [{\citenamefont {Christodoulides}\ and\ \citenamefont
  {Joseph}(1989)}]{PhysRevLett.62.1746}%
  \BibitemOpen
  \bibfield  {author} {\bibinfo {author} {\bibfnamefont {D.~N.}\ \bibnamefont
  {Christodoulides}}\ and\ \bibinfo {author} {\bibfnamefont {R.~I.}\
  \bibnamefont {Joseph}},\ }\bibfield  {title} {\bibinfo {title} {Slow bragg
  solitons in nonlinear periodic structures},\ }\href
  {https://doi.org/10.1103/PhysRevLett.62.1746} {\bibfield  {journal} {\bibinfo
   {journal} {Phys. Rev. Lett.}\ }\textbf {\bibinfo {volume} {62}},\ \bibinfo
  {pages} {1746} (\bibinfo {year} {1989})}\BibitemShut {NoStop}%
\bibitem [{\citenamefont {Jackiw}\ and\ \citenamefont
  {Rebbi}(1976)}]{Jackiw1976}%
  \BibitemOpen
  \bibfield  {author} {\bibinfo {author} {\bibfnamefont {R.}~\bibnamefont
  {Jackiw}}\ and\ \bibinfo {author} {\bibfnamefont {C.}~\bibnamefont {Rebbi}},\
  }\bibfield  {title} {\bibinfo {title} {Solitons with fermion number
  \textonehalf{}},\ }\href {https://doi.org/10.1103/PhysRevD.13.3398}
  {\bibfield  {journal} {\bibinfo  {journal} {Phys. Rev. D}\ }\textbf {\bibinfo
  {volume} {13}},\ \bibinfo {pages} {3398} (\bibinfo {year}
  {1976})}\BibitemShut {NoStop}%
\bibitem [{\citenamefont {Kivshar}(1998)}]{Kivshar1998}%
  \BibitemOpen
  \bibfield  {author} {\bibinfo {author} {\bibfnamefont {Y.~S.}\ \bibnamefont
  {Kivshar}},\ }\bibfield  {title} {\bibinfo {title} {Nonlinear surface modes
  in monoatomic and diatomic lattices},\ }\href
  {https://doi.org/10.1016/S0167-2789(97)00276-5} {\bibfield  {journal}
  {\bibinfo  {journal} {Physica D}\ }\textbf {\bibinfo {volume} {113}},\
  \bibinfo {pages} {248} (\bibinfo {year} {1998})}\BibitemShut {NoStop}%
\bibitem [{\citenamefont {Jayaprakash}\ \emph {et~al.}(2011)\citenamefont
  {Jayaprakash}, \citenamefont {Starosvetsky},\ and\ \citenamefont
  {Vakakis}}]{jayaprakash}%
  \BibitemOpen
  \bibfield  {author} {\bibinfo {author} {\bibfnamefont {K.~R.}\ \bibnamefont
  {Jayaprakash}}, \bibinfo {author} {\bibfnamefont {Y.}~\bibnamefont
  {Starosvetsky}},\ and\ \bibinfo {author} {\bibfnamefont {A.~F.}\ \bibnamefont
  {Vakakis}},\ }\bibfield  {title} {\bibinfo {title} {New family of solitary
  waves in granular dimer chains with no precompression},\ }\href
  {https://doi.org/10.1103/PhysRevE.83.036606} {\bibfield  {journal} {\bibinfo
  {journal} {Phys. Rev. E}\ }\textbf {\bibinfo {volume} {83}},\ \bibinfo
  {pages} {036606} (\bibinfo {year} {2011})}\BibitemShut {NoStop}%
\bibitem [{\citenamefont {Barashenkov}(1996)}]{PhysRevLett.77.1193}%
  \BibitemOpen
  \bibfield  {author} {\bibinfo {author} {\bibfnamefont {I.~V.}\ \bibnamefont
  {Barashenkov}},\ }\bibfield  {title} {\bibinfo {title} {Stability criterion
  for dark solitons},\ }\href {https://doi.org/10.1103/PhysRevLett.77.1193}
  {\bibfield  {journal} {\bibinfo  {journal} {Phys. Rev. Lett.}\ }\textbf
  {\bibinfo {volume} {77}},\ \bibinfo {pages} {1193} (\bibinfo {year}
  {1996})}\BibitemShut {NoStop}%
\bibitem [{\citenamefont {Cuevas-Maraver}\ \emph {et~al.}(2016)\citenamefont
  {Cuevas-Maraver}, \citenamefont {Kevrekidis}, \citenamefont {Saxena},
  \citenamefont {Comech},\ and\ \citenamefont {Lan}}]{PhysRevLett.116.214101}%
  \BibitemOpen
  \bibfield  {author} {\bibinfo {author} {\bibfnamefont {J.}~\bibnamefont
  {Cuevas-Maraver}}, \bibinfo {author} {\bibfnamefont {P.~G.}\ \bibnamefont
  {Kevrekidis}}, \bibinfo {author} {\bibfnamefont {A.}~\bibnamefont {Saxena}},
  \bibinfo {author} {\bibfnamefont {A.}~\bibnamefont {Comech}},\ and\ \bibinfo
  {author} {\bibfnamefont {R.}~\bibnamefont {Lan}},\ }\bibfield  {title}
  {\bibinfo {title} {Stability of solitary waves and vortices in a 2d nonlinear
  dirac model},\ }\href {https://doi.org/10.1103/PhysRevLett.116.214101}
  {\bibfield  {journal} {\bibinfo  {journal} {Phys. Rev. Lett.}\ }\textbf
  {\bibinfo {volume} {116}},\ \bibinfo {pages} {214101} (\bibinfo {year}
  {2016})}\BibitemShut {NoStop}%
\end{thebibliography}%

\end{document}